\documentclass[12pt,a4paper,english]{article}
\pdfoutput=1

\usepackage[bindingoffset=0.3cm,textheight=22.5cm,hdivide={2.7cm,*,2.7cm}, vdivide={*,22cm,*}]{geometry}
\usepackage[bookmarksnumbered=true,breaklinks=true]{hyperref}

\usepackage{amsmath,amsfonts,amssymb,babel,slashed,graphicx,color}

\usepackage[numbers,square,comma,sort&compress]{natbib}

\IfFileExists{dsfont.sty}
	{\usepackage{dsfont}
         \newcommand{\id}{\mathds{1}}}
	{\typeout{Package dsfont.sty was not found, using alternative macros.}
         \let\mathds=\mathbb
         \newcommand{\id}{\mbox{1 \kern-.59em {\rm l}}}}
\let\one=\id

\makeatletter

         \@addtoreset{equation}{section}
\makeatother

\newcommand{\nocontentsline}[3]{}
\newcommand{\tocless}[3]{\bgroup\let\addcontentsline=\nocontentsline#1{#2}#3\egroup}

\newcommand{\qed}{\nobreak \ifvmode \relax \else
      \ifdim\lastskip<1.5em \hskip-\lastskip
      \hskip1.5em plus0em minus0.5em \fi \nobreak
      \vrule height0.75em width0.5em depth0.25em\fi}

\newcommand{\be}{\begin{equation}}
\newcommand{\ee}{\end{equation}}
\newcommand{\eq}[1]{(\ref{#1})}

\def\nn{\nonumber}
\def\bea{\begin{eqnarray}}
\def\eea{\end{eqnarray}}
\def\obar{\overline}

\def\beqa{\begin{eqnarray}} 
\def\eeqa{\end{eqnarray}} 
\def\beq{\begin{equation}} 
\def\eeq{\end{equation}}

\def\a{\alpha}          
\def\b{\beta}           
\def\d{\delta}    
\def\e{\epsilon}                
        
\def\g{\gamma}

\def\l{\lambda} \def\L{\Lambda}

\def\s{\sigma}  
\def\t{\tau}

\def\cA{{\cal A}}  \def\cC{{\cal C}}
\def\cD{{\cal D}}  
 \def\cH{{\cal H}} \def\cI{{\cal I}}
  \def\cL{{\cal L}}
 \def\cN{{\cal N}} \def\cO{{\cal O}}

\def\cV{{\cal V}} \def\cW{{\cal W}}

\newcommand{\R}{\mathds{R}}
\newcommand{\C}{\mathds{C}}

\newcommand{\Z}{\mathds{Z}}

\newcommand{\msu}{\mathfrak{s}\mathfrak{u}}

\newcommand{\mmu}{\mathfrak{u}}

\def\bit{\begin{itemize}}
\def\eit{\end{itemize}}

\def\({\left(}
\def\){\right)}
\def\diag{\mbox{diag}}

\def\del{\partial}

\newcommand{\tr}{\mbox{tr}}

\def\End{{\rm End}}
\def\Hom{{\rm Hom}}

\newcommand{\Di}{{\slashed{D}}}

\DeclareMathOperator{\Tr}{Tr}

 \allowdisplaybreaks[3]

\begin{document}

\renewcommand{\title}[1]{\vspace{10mm}\noindent{\Large{\bf
#1}}\vspace{8mm}} \newcommand{\authors}[1]{\noindent{\large
#1}\vspace{5mm}} \newcommand{\address}[1]{{\itshape #1\vspace{2mm}}}


\begin{flushright}
UWThPh-2015-7 
\end{flushright}

\begin{center}

\title{ \Large Chiral low-energy physics from squashed branes in \\[1ex]
deformed ${\cal N}=4$ SYM}

\vskip 3mm

\authors{Harold C. Steinacker{\footnote{harold.steinacker@univie.ac.at}}
}
 
\vskip 3mm

 \address{ 

{\it Faculty of Physics, University of Vienna\\
Boltzmanngasse 5, A-1090 Vienna, Austria  }  
  }

\vskip 1.4cm

\textbf{Abstract}

\end{center}

 We discuss the low-energy physics which arises on stacks of 
  squashed brane solutions of $SU(N)$ ${\cal N}=4$ SYM, deformed by a cubic soft SUSY breaking potential. 
  A brane configuration is found which leads to a low-energy physics similar to
 the standard model in the broken phase, assuming suitable VEV's of the scalar zero modes.
 Due to the triple self-intersection of the branes, the matter content includes that 
 of the MSSM with precisely 3 generations and right-handed neutrinos. 
 No exotic quantum numbers arise, however there are extra chiral superfields
 with the quantum numbers of the Higgs doublets, the $W,Z$, $e_R$ and $u_R$, whose 
 fate depends on the details of the rich Higgs sector.
 The chiral low-energy sector is complemented by a heavy mirror sector with the opposite chiralities,
 as well as super-massive Kaluza-Klein towers completing the $\cN=4$ multiplets.
 The sectors are protected by two gauged global $U(1)$ symmetries.

\vskip 1.4cm

\textbf{keywords:}  fuzzy extra dimensions; $N=4$ super-Yang-Mills;
mirror fermions; chiral gauge theory

\newpage

\tableofcontents

\section{Introduction}\label{sec:background}

$\cN=4$ Super-Yang-Mills (SYM) is the most (super)symmetric of all 4-dimensional field 
theories without gravity.
As such it has played a prominent role ever since its discovery,
even though it is usually considered as ``too round'' for real physics.
More structure can be introduced by considering deformations of that model, notably 
by adding soft SUSY breaking terms to the potential. Then interesting patterns 
of spontaneous symmetry breaking can occur, inducing even more structure at low energy.
A well-known example is the generation of fuzzy spheres, realized by the 
vacuum expectation values (VEV's) of the 
matrix-valued scalar fields. Due to the Higgs effect, the model then behaves like a higher-dimensional 
model on $\R^4\times S^2_N$ \cite{DuboisViolette:1989at,Myers:1999ps,Iso:2001mg,Polchinski:2000uf,
Berenstein:2002jq,Aschieri:2006uw,Andrews:2006aw,Steinacker:2007ay}. 
Recently, a much richer class of such solutions was found \cite{Steinacker:2014lma,Steinacker:2014eua}
 in the presence of a 
cubic SUSY-breaking potential corresponding to a holomorphic 3-form. These solutions
can be interpreted as projected or ``squashed'' fuzzy coadjoint orbits $\cC[\mu]$  of $SU(3)$.
Due to their self-intersecting geometry, they lead to 3 generations of massless 
fermions and scalar fields.

In this paper, we discuss these squashed brane solutions  in more detail, 
and study some aspects of the resulting low-energy physics on stacks of such branes.
Since there are  massless scalar fields, it is natural 
(due to the presence of cubic interactions) to assume that some of them take non-trivial VEV's.
The main point to be emphasized here is that for suitable such VEV's,
the resulting low-energy sector behaves like a chiral gauge theory,
in the sense that different chiralities of the fermionic (would-be) zero modes couple differently to the 
spontaneously broken massive gauge fields.
Since this is a fundamental property of the standard model,
that class of models becomes quite interesting for physics.


We first review and re-derive the fermionic and bosonic zero modes from a field-theory perspective,
recovering results in \cite{Steinacker:2014lma}. The approach given here is based on 
two global symmetries $U(1)_{K_i}$ which are respected by the background up to gauge transformations;
these allow a coherent treatment of the bosonic and fermionic modes, and are very useful 
in understanding the interactions. 
The ``regular'' zero modes on a stack of $\cC[\mu_i]$ branes can be organized in terms of a quiver,
with 3 families of chiral superfields transforming in the bi-fundamental of 
gauge group $U(n_i)$ arising on the coincident branes.
They have specific $U(1)_{K_i}$ weights in the $\msu(3)$ weight lattice.
Gauge fields and gauginos arise in vector supermultiplets. 
Nevertheless, the low-energy theory is not supersymmetric.  
These massless scalar modes will be dubbed ``Higgs'' modes henceforth.

Without attempting a full understanding of the rich Higgs sector in this paper, we consider
some of the possible Higgs configurations, and elaborate the resulting physics in some detail using the new tools.
In particular, we give a brane configuration which 
leads to  the correct pattern of leptons and quarks coupling to the gauge fields 
of the standard model in its broken phase. 
This leads to an extension of the MSSM, where each chiral super-multiplet
has an extra mirror copy with the opposite chirality, which acquires a higher (by assumption)
mass from the mirror Higgs. 
The present scenario\footnote{This is basically a refinement of an outline given in 
\cite{Steinacker:2014eua}, in a slightly different more conservative setting.}  
improves upon the analogous background solutions  in
\cite{Steinacker:2014fja} and related proposals \cite{Aoki:2014cya}
in several ways. First and foremost, there are necessarily 3 generations
due to the triple self-interacting geometries, resulting in a $\Z_3$ family symmetry
(which may subsequently be broken). Moreover, the chiral
low-energy sector is protected from recombining with the massive mirror 
fermions due to two exact $U(1)_{K_i}$ symmetries.
These are combinations of the $R$-symmetry and the gauge symmetry, which are preserved by the background.
In this way, a stable chiral low-energy physics can arise from the underlying non-chiral 
$\cN=4$ theory. Furthermore, the scale of the mirror fermions can in principle be much higher than the electroweak scale, 
for large branes.

The present scenario is somewhat reminiscent of higher-dimensional string-theoretical 
(and field-theoretical) 
 models such as \cite{Aldazabal:2000cn,Manousselis:2004xd}. However 
it is much more radical and simple, since the chiral low-energy behavior is not 
put in by hand but arises from spontaneous symmetry breaking.
Even if it may seem unlikely that such a scenario could be realistic, 
it is certainly worthwhile to explore the possible scope of these deformed $\cN=4$ models,
given their special status in field theory.

Due to the complicated Higgs sector, no attempt is made in this paper
to find the minima and to justify the assumed Higgs configuration.
However, the basic result that certain Higgs configurations lead to chiral low-energy
sector and a massive mirror sector is fully justified, and verified  numerically.
Also, the structure of leptons and quarks is very clear and convincing.
However there is a rather complicated sector of modes with the 
quantum numbers of the two Higgs doublets and the electroweak gauge bosons, which 
is not worked out in detail.
Some numerical computations are  performed to gain some more insights, however this 
clearly requires more detailed investigations.

This paper is intended to be largely self-contained, and written from a field-theory perspective.
Rather than just relying on the previous papers \cite{Steinacker:2014lma,Steinacker:2014eua},
the necessary results are re-derived in a more transparent 
way, emphasizing the role of the two $U(1)_{K_i}$ symmetries.
Although this increases the length,  the paper should be more accessible in this way.

\section{Squashed $SU(3)$ branes in deformed $\cN=4$ SYM}

We start with the action of $\cN=4$ $SU(N)$ SYM, which is organized most transparently 
in terms of 10-dimensional SYM reduced to 4 dimensions:
\begin{align}
 S_{\rm YM}
&=  \int d^4 x \ \frac 1{4g_{N}^2} \tr\Big(-F^{\mu\nu} F_{\mu\nu} 
 - 2  D^\mu \Phi^a D_\mu \Phi_a +  [\Phi^a,\Phi^b][\Phi_a,\Phi_b] \Big)\nn\\
&\qquad \qquad +  \tr\Big(\bar\Psi\g^{\mu} i D_\mu \Psi + \bar\Psi\Gamma^a [\Phi_a,\Psi]\Big) .
\label{N=4SYM}
\end{align}
Here $F_{\mu\nu}$ is the field strength, $D_\mu = \del_\mu - i [A_\mu,.]$ the covariant derivative,
$\Phi^a,\ a \in \{1,2,4,5,6,7\}$ are 6 scalar fields, 
$\Psi$ is a matrix-valued Majorana-Weyl (MW) spinor of $SO(9,1)$ dimensionally  
reduced to 4-dimensions, and $\Gamma^a$ arise from the 10-dimensional gamma matrices.
 All fields transform take values in $\mmu(N)$ and transform
 in the adjoint of the $SU(N)$ gauge symmetry. The global $SO(6)_R$ symmetry is manifest.
It will be useful to work with dimensionless scalar fields labeled by 
the six roots\footnote{Here we use field theory conventions, while in \cite{Steinacker:2014lma}  
group-theory friendly conventions are used. In particular, 
the $\a_i$ are related to  the standard 
basis $\tilde\a_i$ of positive roots of $\msu(3)$ used in group theory via 
$\a_1=\tilde\a_1,\a_2=\tilde\a_2,\a_3=-\tilde\a_3$, such that 
$\a_1+\a_2+\a_3=0$; this is more natural here.} $\pm\a_i$ of $\msu(3)$, 
\begin{align}
 \Phi_\a =   m X_\a, \qquad \a\in \cI =  \{\pm\a_i,\ i=1,2,3 \} \ \subset \R^2
\end{align}
 with $\a_1+\a_2+\a_3=0$.
These $\a\in \cI$
 are viewed as points in $\R^2$ forming a regular hexagon 
 (see figure \ref{fig:spin-states-roots-beta}), with corresponding 
 Weyl chambers defined by the Weyl group $\cW$ of reflections along these roots.
Here $m$ has the dimension of a mass.
Explicitly,
\begin{align}
 X_1^\pm &= \frac 1{\sqrt{2}}(X_4\pm i X_5) \equiv X_{\pm \a_1}, \nn\\
 X_2^\pm &= \frac 1{\sqrt{2}}(X_6\mp i X_7)  \equiv X_{\pm \a_2}, \nn\\ 
 X_3^\pm &= \frac 1{\sqrt{2}}(X_1\mp i X_2)  \equiv X_{\pm \a_3} \ 
 \label{X-T-definition}
\end{align}
To introduce a scale and to allow non-trivial ``brane'' solutions, we add  soft terms to the potential,
\begin{align}
 \cV[\Phi] &=  \frac {m^4}{g_N^2}  \big(V_4[X] + V_{\rm soft}[X]\big)
 \end{align}
 where
 \begin{align}
 V_4[X] &= -\frac 14 \tr \sum_{\a,\b\in\cI}[X_\a,X_\b][X^\a,X^\b] , \nn\\
 V_{\rm soft}[X] &= 4\, \tr \big(-[X_1^+,X_2^+] X_3^+ - [X_2^-,X_1^-] X_3^- + M_{i}^2 X_i^- X_i^+ \big) 
 \label{V-soft}
\end{align}
thereby fixing the scale $m$.
The cubic potential can be written as  
 \begin{align}
 V_3(X) = - \frac {4}3\tr \big(\varepsilon_{ijk} X_i^+ X_j^+ X_k^+  + h.c.\big)\ ,
\label{cubic-flux}
\end{align}
corresponding to a holomorphic 3-form on $\R^6$.  
Rewritten in a  real basis, this is recognized as the structure constants of $\msu(3)$
projected to the root generators \cite{Steinacker:2014lma}.

We will mostly set $M_{i} = 0$ in this paper. 
Then SUSY is explicitly broken, and the global $SO(6)_R$ symmetry is broken to $SU(3)_R$ by the cubic term.
However as show in appendix \ref{sec:N=1}, some supersymmetry can be preserved
for a suitable choice of $M_{i}$ (and corresponding fermionic  terms).
More precisely, there is a specific
$\cN=1^*$ deformation of $\cN=4$ SYM \cite{Karch:1999pv,Argyres:1999xu,Polchinski:2000uf}
with potential \eq{V-soft}.
However this requires $M_i$ to be outside of the domain which admits the squashed brane solutions
of interest here. Nevertheless, this observation should help to understand  the quantum corrections of the model,
which is left for future work. Here we focus on the classical aspects of the model.

\paragraph{Perturbation of the background.}

Let us add a  perturbation $\phi^\a$ to the background $X^\a$,
\begin{align}
\Phi^\a  = m(X^\a + \phi^\a) .
\end{align}
This will lead to further symmetry breaking and interesting low-energy physics
in the zero-mode sector of the background $X$.
The complete potential is easily worked out,
\begin{align}
  V(X+\phi) 
&= V(X) +   \tr\Big(\phi^\a \Box_X X_\a + X^\a \Box_\phi \phi_\a 
        +\frac 12\phi^\a  \big(\Box_X + 2 \Di_{ad} \big) \phi_\a - \frac 12 f^2 \Big) \nn\\
     &\ \ + 4\tr \big(- \varepsilon_{ijk} \phi_i^+ X_j^+ X_k^+ 
      - \varepsilon_{ijk} \phi_i^+ \phi_j^+ X_k^+  + M_{i}^2 \phi_i^- X_i^+ 
      + \frac 12 M_{i}^2 \phi_i^- \phi_i^+ + h.c. \big).
 \label{V-full-expand}
\end{align}
Here 
\begin{align}
 f &= i[\phi_\a,X^\a]  
\end{align}
can be viewed as gauge-fixing function in extra dimensions, and we define
\begin{align}
 \Box_X &= \sum_{a\in\cI} [X_\a,[X^\a,.]]
 = [X_j^+, [X_j^-,.]] + [X_j^-, [X_j^+,.]] ,  \\
(\Di_{ad} \phi)_\a &= \sum_\b [[X_\a,X^{\b}],\phi_{\b}] 
=  ((\slashed{D}_{\rm mix}  + \slashed{D}_{\rm diag}) \phi)_\a \nn\\
(\slashed{D}_{\rm mix} \phi)_\a &= \sum_{\b\neq \a} [[X_\a,X^{\b}],\phi_{\b}]  \nn\\
(\slashed{D}_{\rm diag} \phi)_\a &= [[X_\a,X_{-\a}],\phi_{\a}]  \qquad\mbox{(no sum)}
 \label{higgs-decouple-1}
\end{align}
following \cite{Steinacker:2014eua},
noting that 
\begin{align}
 X^\a &=  X_{-\a}  .
\end{align}
In particular, the equations of motion (eom) for the background $X$ can be written as
\begin{align}
 0=\big(\Box_4  + m^2(\Box_X + 4 M_{i}^2)\big)X_i^+ + 4 m^2 \varepsilon_{ijk} X_j^- X_k^- 
 \label{eom-roots}
\end{align}
where $\Box_4 = -D_\mu D^\mu$ is the 4-dimensional covariant d'Alembertian.

\subsection{Squashed brane solutions}
\label{sec:solutions}

\begin{figure}
\begin{center}
 \includegraphics[width=0.35\textwidth]{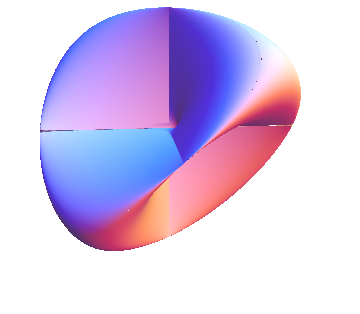}
 \end{center}
 \caption{3-dimensional section of squashed $\C P^2$, taken from \cite{Steinacker:2014lma}.}
 \label{fig:squashedCP2}
\end{figure}

It is well-known that the above potential has fuzzy sphere solutions $X_i^\pm \sim c_i^\pm J_i$ 
where $J_i$ are generators of $\msu(2)$
\cite{DuboisViolette:1989at,Vafa:1994tf,Myers:1999ps,Polchinski:2000uf,Iso:2001mg,Berenstein:2002jq,Andrews:2006aw,Aschieri:2006uw}.
However as shown in \cite{Steinacker:2014lma}, there are also solutions with much richer structure
corresponding to (stacks of) squashed fuzzy coadjoint $SU(3)$ orbits $\cC_N[\mu]$, obtained by 
the following ansatz 
\begin{align}\fbox{$ \ 
 X_i^\pm = r_i \pi(T_i^\pm)
  \ $} \ .
  \label{basic-branes}
\end{align}
Here
\begin{align}
 T_1^\pm &\equiv T_{\pm \a_1}, \qquad T_2^\pm \equiv T_{\pm \a_2},  \qquad
 T_3^\pm \equiv T_{\pm \a_3} 
 \label{X-T-definition-roots}
\end{align}
 are {\em root} generators of $\msu(3)_X$,
$\pi$ is any representation on $\cH \cong \C^N$, and
$\a_{1}, \a_2$ are the simple roots 
with $\a_3 = -(\a_1+\a_2)$. 
In these conventions, the Lie algebra relations are
\begin{align}
 [T_\a,T_\b] &= \pm T_{\a+\b}, \qquad 0 \neq \a+\b\in \cI  \nn\\
 [T_{\a_i},T_{-\a_i}] &= H_{i} \equiv H_{\a_i} \nn\\
 [H,T_\a] &= \a(H) T_\a \ , 
 \label{Cartan-Weyl}
\end{align}
with
$[T_1^+,T_2^+] = T_3^-$ and $\a_i(H_i) = (\a_i,\a_i) =2$ where 
$(.,.)$ denotes the Killing form of $\msu(3)$.
Using these Lie algebra relations, the equations of motion \eq{eom-roots} become 
\begin{align}
0  
 &=  r_1\big(2 r_1^2 +r_2^2 + r_3^2 - 4 \frac{r_2 r_3}{r_1} + 4M_{1}^2 \big) T_1^+ \nn\\
 0 
 &= r_2\big(r_1^2 + 2 r_2^2 + r_3^2 - 4 \frac{r_1 r_3}{r_2}  + 4M_{2}^2\big) T_2^+  \nn\\
 0 
 &= r_3\big(r_1^2 + r_2^2 + 2r_3^2 - 4 \frac{r_1 r_2}{r_3}  + 4M_{3}^2\big) T_3^+ .
  \label{eom-general}
\end{align}
Assuming  $M_i=0$ for simplicity, these equations have 
the non-trivial solution
\begin{align}
 r_i = 1 \equiv r \ .
\end{align}
For $\pi=\pi_\mu$ an irreducible representation (irrep) with highest weight $\mu$ acting on $\cH_\mu$,
these solutions can be interpreted as quantized or fuzzy 
coadjoint orbits $\cC[\mu] \subset\msu(3)_X \cong \R^8$ 
projected to $\R^6$ along two Cartan generators \cite{Steinacker:2014lma}.
Generically these are 6-dimensional (fuzzy) varieties, while for $\mu =(n,0)$ and $\mu = (0,n)$ they are 
4-dimensional projections of (fuzzy) $\C P^2$.
Here $\mu=(n_1,n_2)$ denotes the Dynkin labels of $\mu$. 
Such a ``squashed'' $\C P^2$ has a triple self-intersection  at the origin, as  visualized in figure 
\ref{fig:squashedCP2}.
We will see that pairs of fermionic zero modes 
arise at the intersections, connecting the different sheets. 

To organize the degrees of freedom, we note that 
these solutions defines an embedding $SU(3)_X \subset SU(N)$, which acts via the adjoint on all the fields
in the theory.  Decomposing the $\msu(N)$-valued fields into harmonics i.e. irreps of this $SU(3)_X$
\begin{align}
\msu(N) \cong \End(\cH) = \oplus_\L  n_\L \cH_\L 
\end{align}
(here $\cH_\L$ denotes the highest weight irreps)
allows to understand the physics of the fluctuations on such a background,
even though the $SU(N)$ gauge symmetry is broken completely
for irreducible $\pi_\mu$. In particular,
the $SU(3)_X$ gauge transformations act on the scalar fields as
\begin{align}
 X_\a \to \pi(g) \pi(T_\a) \pi(g^{-1}) = \L(g)_\a^\b \pi(T_\b) +  \L(g)_\a^i \pi(H_i) 
 \label{scalars-trafo-general}
\end{align}
(here $\L_\a^\b$ is the $8$-dimensional representation of $SU(3)_X$).
Now restrict to the Cartan subalgebra or the torus
$U(1)\times U(1) \subset SU(3)_X$, which is sufficient 
to specify the weights in the various $\msu(3)_X$ harmonics.
Then the last term in \eq{scalars-trafo-general} vanishes, and the six scalar fields $X_\a$ 
transform linearly, corresponding to the six  non-zero weights in $(1,1)$ of $\msu(3)_X$.
This organization will be very useful.

The potential has a global $SU(3)_R\subset SU(4)_R$ symmetry,
which is broken to $SU(2) \times U(1)$ or $U(1)^2$ in the presence of masses $M_i\neq 0$.
We denote with $\t_i$ the traceless $U(1)_i \subset U(3)_R$ generator which has eigenvalue $1$ on $X_i^+$ 
and $-\frac 12$ on the $X_j^+$ with $j\neq i$, or more formally 
\begin{align}
 2\t_i\phi_\a = (\a_i,\a) \phi_\a \ .
\end{align}
Then $\sum_i \t_i = 0$, and the action of $2\t_i$ on the scalar fields coincides with the 
adjoint action of the Cartan generators $H_{\a_i}$ of $\msu(3)_X$.
In other words, the background $X^\a$ is annihilated by the following generators
\begin{align}
K_i := 2\t_i - [H_{\a_i},.] ,  \qquad i=1,2,3  
\label{combined-U1-symm}
\end{align}
which satisfy $K_1+K_2+K_3=0$, and generate a $U(1)_K \times U(1)_K$ symmetry of the background.
Their charges 
are obtained by adding the 
(rescaled and rotated, cf. figure \ref{fig:6-parameters}) $(1,0) + (0,1)$ weights
of $\msu(3)_R$ to the non-zero weights of $(1,1)$  of $\msu(3)_X$.
In particular, the charges of $U(1)_i^K$ are points in the weight lattice of $\msu(3)_X$.
This will be very important to characterize and protect the zero modes.

Now we can understand the Goldstone bosons arising from the broken global symmetries.
The  background breaks the global $SU(3)_R$ symmetry, but the traceless 
$U(1)_i$ with generators $\t_i$ are equivalent to gauge transformations (i.e. ``gauged``).
Therefore there will be only $8-2$ physical Goldstone bosons, as
the two $U(1)_{\t_i}$ modes are eaten by the massive gauge bosons.
These 6 physical Goldstone bosons are identified in appendix \ref{sec:goldstone}
with the 6 exceptional zero modes 
in the $(1,1)\subset \End(\cH)$ as discussed below. 

Finally, the background admits a  $\Z_3$ symmetry, which cyclically permutes the 
$X_i^\pm$. This is part of the $SU(3)_R$ symmetry, and also part of the 
Weyl group\footnote{The potential is in fact invariant under the full
Weyl group $S_3$ of $SU(3)_X$.} of $SU(3)_X$.
It is also interesting to recall that the global $SU(4)_R$ is anomalous, and there is an
associated Wess-Zumino term \cite{Tseytlin:1999tp}.
This might be important for the effective description 
of the 6 physical Goldstone bosons.

\subsection{Scalar zero modes on squashed branes}
\label{sec:scalar-zeromodes}

Let  $M_i=0$ from now on.
Then the squashed brane backgrounds $X_\a$ admit a number 
of zero modes  $\phi^{(0)}_\a$.
To see this, we note that the bilinear form defined by $\slashed{D}_{\rm mix}$ 
on a background \eq{basic-branes} can be simplified e.g. as follows
\begin{align}
 \tr \big(\phi_i^-(\slashed{D}_{\rm mix} \phi)^+_i\big) 
  &= \sum_{j\neq i} \tr (\phi_i^-[[X_i^+,X_{j}^+],\phi_{j}^-]) 
  =  - \varepsilon_{ikj} \tr(\phi_i^-[X_{k}^-,\phi_{j}^-])
\end{align}
using $[T_i^+,T_j^+] = \varepsilon_{ijk} T_k^-$ and $r=1$.
This has precisely the form of the quadratic contribution from the cubic potential 
 \eq{V-full-expand}. Therefore the quadratic terms in the potential can be written as
\begin{align}
 V_2[\phi] &= \tr \phi^\a \cO_V \phi_\a  \ ,   
  \qquad \cO_V = \Box_X + 2\slashed{D}_{\rm diag} - 2\slashed{D}_{\rm mix} \ .
 \label{fluct-equation}
\end{align}
It was shown in \cite{Steinacker:2014lma} that $\cO_V$
is positive semi-definite for all representations $\pi$.
The zero modes of $\cO_V$ fall into two classes, denoted as regular and exceptional
zero modes. Let us first focus on the regular zero modes, which are given by 
solutions of the  decoupling condition \cite{Steinacker:2014lma,Steinacker:2014eua}
\begin{align}
 \Di_{\rm mix} \phi_\a^{(0)} = 0 .
 \label{D-mix-vanish}
\end{align}
Here we shall provide a group-theoretic characterization 
of the regular zero modes, which implies \eq{D-mix-vanish}; it is then straightforward to 
show that they are zero modes.
Recall that the background respects the $U(1)$ generators\footnote{This symmetry was also used in 
\cite{Steinacker:2014eua} to classify excitations on spinning brane backgrounds.} 
$K_i = 2\t_i-[H_{\a_i},.]$ \eq{combined-U1-symm}, and
consider  the ''$\t$-parity`` generator $\t$ in $U(3)_R$ defined by
\begin{align}
 \t \phi_i^\pm = \pm \phi_i^\pm \ ,
 \label{tau-parity} 
\end{align}
which is broken by the cubic potential. Then
\begin{align}
 \t \Di_{\rm mix} &= - \Di_{\rm mix} \t \nn\\[1ex]
 \Di_{\rm mix} K_i &=  K_i \Di_{\rm mix} \ .
 \label{Dmix-tau}
\end{align}
Now  fix some highest weight module $\cH_{\L} \subset \End(\cH)$, and 
consider the set of  $U(1)_{K_i}$ weights of $\phi_\a \in \cH_{\L}$, 
given by the 6 nonzero weights $\a \in (1,1)$ of $\msu(3)_X$ minus
the weights in $\cH_\L$. 
Among these, consider the 6 {\em extremal weights}\footnote{These are the corners of the 
convex set of weights in $\R^2$,
or equivalently of the maximal irrep in $(1,1)\otimes \cH_{\L^+}$.
The conventions differ in an inessential way from 
the ones in \cite{Steinacker:2014eua}.} $\L'$, and denote the corresponding modes as 
\begin{align}
 \phi_{\a,\L'}^{(0)} &= Y_{\l}\ , \qquad \L'=\a-\l, \qquad  Y_{\l} \in \cH_{\L} \ . 
\end{align}
Here $Y_{\l}$ is an extremal weight vector with weight $\l$ in  $\cH_{\L}$.
These $\phi_{\a,\L'}^{(0)}$ have charge 
$\L'= \a-\l$ under the $K_i$, 
corresponding to a point of the $\msu(3)_X$ weight lattice in 
(the interior of) the Weyl chamber of $\a$. These are the {\em regular zero modes}. 
They have eigenvalue  $\t=\pm 1$ determined by the parity of the Weyl chamber of $\a=\pm\a_i$.
Since there is only one such state for any such (extremal!) $\L'$
and $\Di_{\rm mix}$ preserves $\L'$, \eq{Dmix-tau} implies that  
\begin{align}
 \Di_{\rm mix} \phi^{(0)}_{\a,\L'} = 0 \  .
 \label{bosonic-zeromodes-Dmix}
\end{align}
Using the extremal weight property, it 
is then easy to verify that these are zero modes
\begin{align}
 \cO_V \phi^{(0)}_{\a,\L'} =  (\Box_X + 2\slashed{D}_{\rm diag})\phi^{(0)}_{\a,\L'} =  0 \ ;
 \label{bosonic-zeromodes}
\end{align}
e.g. for $\a=\a_3$, we have (cf.\cite{Steinacker:2014eua})
\begin{align}
 \Box_X \phi^{(0)}_{\a_3,\L'} &=  r^2 ([T_1^+,[T_1^-,.]] + [T_2^+,[T_2^-,.]]+[T_3^-,[T_3^+,.]]) 
 \phi^{(0)}_{\a_3,\L'} \nn\\
 &=  r^2 ([H_1,.] + [H_2,,.] - [H_3,.])  \phi^{(0)}_{\a_3,\L'} \ , \nn\\[1ex]
 \slashed{D}_{\rm diag}\phi^{(0)}_{\a_3,\L'} &= r^2 [[T_{3},T_{-3}],\phi_{\a_3}^{(0)}] 
 = r^2 [H_3,\phi_{\a_3}^{(0)}] \ 
\end{align}
hence \eq{bosonic-zeromodes} follows from $H_1+H_2+H_3=0$.
We will find superpartners of these regular zero modes in section \ref{sec:fermions}.
Particular examples of such modes are given by 
\begin{align}
 \phi_{\a,(n+1)\a}^{(0)} = (T_{-\a})^{n} .
 \label{zeromodes-examples}
\end{align}
Observe that they have eigenvalue $\t=\pm 1$ determined by the Weyl chamber of $\a=\pm\a_i$.
A possible background with such a ``Higgs'' switched on would then be 
\begin{align}
 \Phi_\a = m(X_\a + \varepsilon_\a T_{-\a}^n) .
\end{align}
On  a single  squashed $\C P^2_N$ brane, these 
exhausts all regular zero modes.
Observe that there are 6 such zero modes even for degenerate $\L$
such as $\L=(m,0)$.
Some intuition can be gained by noting that the regular zero modes with maximal 
$\L'$ on squashed $\C P^2_N$ link the 3 intersecting $\R^4$ sheets at the origin, 
with polarization along the common $\R^2$ \cite{Steinacker:2014lma}. 
More generally, the regular zero modes can be interpreted as strings linking these sheets,
shifted along their intersection\footnote{For examples of such string-like modes
see e.g. \cite{Andronache:2015sxa}.}.

For harmonics $\cH_\L\in \End(\cH)$ with $\L = (m-2,1)$ and $\L = (1,m-2)$, 
there are in addition 3 exceptional zero modes 
\cite{Steinacker:2014lma} with $\L'=(m,0)$ resp. $\L' = (0,m)$, 
which have mixed polarizations.
The most important among these arise for $\L=(1,1)$: 
these correspond to the 6 physical 
Goldstone bosons which arise from $SU(3)_R$ as discussed above, see appendix \ref{sec:goldstone}. 
The full set of zero modes for squashed $\C P^2$ branes
can then be obtained from the mode decomposition 
\begin{align}
  (n,0) \otimes (0,m) &= (n,m) + (n-1,m-1) + ... + (n-m,0) \nn\\
 (n,0) \otimes (m,0) &= (n+m,0) + (n+m-2,1) + (n+m-4,2) + ... + (...,m) 
\end{align}
assuming $n \geq m$. 
There is a set of 6 exceptional zero modes in  $(n,0) \otimes (0,n)$ given by the Goldstone 
bosons, and typically 3 exceptional zero modes for $(n,0) \otimes (m,0)$,
or $6$ for $n+m=3$.

From now on, we will collectively denote the set of these
scalar  (``would-be'') zero modes as Higgs sector, anticipating that they may acquire 
some VEV or some mass.

\paragraph{Higgs connecting different branes.}

Now consider backgrounds consisting of several branes,  described by reducible 
representations  in \eq{basic-branes}.
For example, the matrix modes on a stack consisting of 
one $\cC[\mu_L]$ and one $\cC[\mu_R]$ brane decompose as
\begin{align}
 \End(\cH) &=  \End(\cH_{\mu_L}) +  \End(\cH_{\mu_R}) 
  + \cH_{\mu_L}\otimes\cH_{\mu_R}^* + \cH_{\mu_L}\otimes\cH_{\mu_R}^*  .
\end{align}
To be specific, assume that  $\mu_L = (1,0)$ and $\mu_R = (0,1)$.
Then $\End(\cH_{\mu_L}) = (1,1) + (0,0) = \End(\cH_{\mu_R})$,
each leading to 6 regular zero modes with $\L' \in \cW (2,2)$ from $\cH_{(1,1)}$,
6 regular zero modes with $\L' \in \cW (1,1)$ 
from $\cH_{(0,0)}$  corresponding to translations 
in the internal $\R^6$, and 6 exceptional zero modes with $\L'\in\{\cW(3,0), \cW(0,3)\}$
from $\cH_{(1,1)}$  corresponding to 
the $SU(3)_R$ Goldstone bosons.
Here $\cW(n_1,n_2)$ denotes the set of weights given by the 
action of the Weyl group $\cW$ on $\L'=(n_1,n_2)$.
On the other hand, the regular modes connecting different branes can be written as
\begin{align}
 \phi_{\a,\L'}^{(0)} = \varphi_{\a}^{ij}\, |\mu_L^i\rangle\langle \mu_R^j| 
 \ \in  \ \cH_{\mu_L}\otimes\cH_{\mu_R}^*   \cong   \cH_{(2,0)} \oplus \cH_{(0,1)}, 
  \label{zeromodes-bosons-LR}
\end{align}
where $\mu^i_{L,R}$ are  weights in $\cH_{\mu_{L,R}}$, and 
$\L' = \a-(\mu_L^i - \mu_R^j)$ is in the Weyl chamber opposite to $\a$.
This leads to 6 regular zero modes with $\L' \in \cW (3,1)$ from $\cH_{(2,0)}$,
6 regular zero modes with $\L' \in \cW (1,2)$ from $\cH_{(0,1)}$,
and 3 exceptional zero modes with $\L'\in \cW(2,0)$ from $\cH_{(0,1)}$.
 Since the $|\mu_R^i\rangle$ can be viewed as coherent states 
on  $\cC[\mu_R]$ located at the origin \cite{Steinacker:2014lma}, 
 the $\cH_{(0,1)}$ modes 
 are strings linking the sheets of $\cC[(1,0)]$ and $\cC[(0,1)]$ with a 2-dimensional 
intersection at the origin, while the $\cH_{(2,0)}$ modes  
are strings linking coinciding sheets  at the origin.
Finally, the 
modes connecting $\cC[(1,0)]$ with a point brane $\cC[0]$ arise from 
\begin{align}
 \phi_{\a,\L'}^{(0)} = \varphi_{\a}^{i}\, |0\rangle\langle \mu_R^i|  \ \in  \ (0,0)\otimes (1,0) 
  \cong \cH_{(1,0)}\ , 
 \label{zeromodes-bosons-L0}
\end{align}
leading to 6 regular zero modes with $\L' \in \cW (2,1)$,
and 3 exceptional zero modes with $\L' \in \cW (0,2)$.


\section{A standard-model-like brane configuration}
\label{SM-branes}

Now consider a background consisting of two coincident (isomorphic)  branes $\cC[\mu_{L}]_u, \ \cC[\mu_{L}]_d$, 
and two additional (typically different)  branes $\cC[\mu_{Ru}]$ and $\cC[\mu_{Rd}]$.
We assume that the scale $m$ of these branes is very high.
Furthermore we add a ``leptonic'' point brane $\cD_l \cong \cC[0]$, 
and 3  ``baryonic'' point branes $\cD_{b_j}\cong \cC[0], \ j = 1,2,3$.
Hence the matrices of $\cN=4$ SYM act on
\begin{align}
 \cH = \cH_{{L}}^2 \oplus \cH_{{Ru}} \oplus \cH_{{Rd}} \oplus \C \oplus \C^3 .
  \label{Hilbert-decomp}
\end{align}
If realized within $U(N)$ SYM (instead of $SU(N)$), 
this background admits a $U(2)_L \times U(1)_{Ru} \times U(1)_{Rd} \times U(1)_l \times U(3)_c$ 
gauge symmetry, or  $U(2)_L \times U(2)_{R} \times U(1)_l \times U(3)_c$ if $\mu_{Ru} = \mu_{Rd}$.
Such stacks of branes 
might be bound by quantum effects in $\cN=4$ SYM, which are related to supergravity
and typically induce an attractive interaction between branes with different flux 
\cite{Chatzistavrakidis:2011gs,Blaschke:2011qu,Chepelev:1997av,Ishibashi:1996xs}.

Now we switch on some ``Higgs'' links  between these branes, realized by (would-be) zero modes $\phi_\a$
linking some extremal states of the various $\cH_i$ as in \eq{zeromodes-bosons-LR},
\begin{align}
 \phi_\a  = \sum \varphi_{\a}^{ij} |\mu_{L}\rangle_{i} \langle\mu_R|_{j}   + ...
  \ \in  \ \oplus \ \Hom(\cH_{j},\cH_{i}) \ .
\end{align}
First,  assume that the point-brane  $\cD_l$
is linked  to the extremal weight states of $\cC[\mu_{Ru}]$
as in \eq{zeromodes-bosons-L0},
\begin{align}
 \phi_S = \sum  \varphi_S |0\rangle_l\langle\mu_{Ru}|_u 
 \label{s-Higgs}
\end{align}
dropping the polarization indices.
Assuming that the scale of $\varphi_S$ is high\footnote{We might as well 
consider $\cD_l\cup_{\phi_S}\cC[\mu_{Ru}]$ as a single brane.},
the unbroken gauge symmetry is reduced to
 $U(2)_L \times U(1)_{Ru,l} \times U(1)_{Rd} \times U(3)_c$,
 where $U(1)_{Ru,l}$ is generated by $\one_l + \one_{Ru}$.
To write this in a more suggestive form, we  introduce the hypercharge generator
\begin{align}
  Y &:= \one_{Ru} -\one_{Rd} + L - B \ ,
\label{Y-def}
 \end{align} 
and
\begin{align}
 T_5 &= B-L+\Xi 
\end{align}
 (as in \cite{Steinacker:2014fja}), where
 \begin{align}
  \Xi &=\one_{Lu } + \one_{Ld} - (\one_{Rd} + \one_{Ru}) ,  \nn\\
    B &= \frac 13 \one_b, \nn\\
    L &= \one_l .
 \label{t-5-generator}
\end{align}
 $\Xi$ will act as chirality $\gamma_5$ in the light sector due to \eq{chiral-light-4D}.
Then the unbroken gauge symmetry can be written as 
$SU(3)_c \times SU(2)_L \times U(1)_Y \times U(1)_B \times U(1)_5 \times U(1)_{\rm tr}$,
where $U(1)_5$ and $U(1)_B$  will be anomalous in the light sector but not in the full model, and 
$U(1)_{\rm tr}$ is the trace-$U(1)$. The latter can be dropped in $\cN=4$ SYM, 
but acquires an interesting role related to gravity 
in the IKKT matrix model \cite{Steinacker:2010rh}. 
This leaves exactly the gauge group of the standard model 
$SU(3)_c \times SU(2)_L \times U(1)_Y$, 
extended by  $U(1)_B, U(1)_5$, 
and possibly $U(1)_{\rm tr}$. 
The $U(1)_5$ will also be broken by the electroweak Higgs as elaborated below.
All other gauge bosons are massive with mass set by the scale $m$ or $\varphi_S$; 
we will ignore these from now on.

\begin{figure}
\begin{center}
 \includegraphics[width=0.55\textwidth]{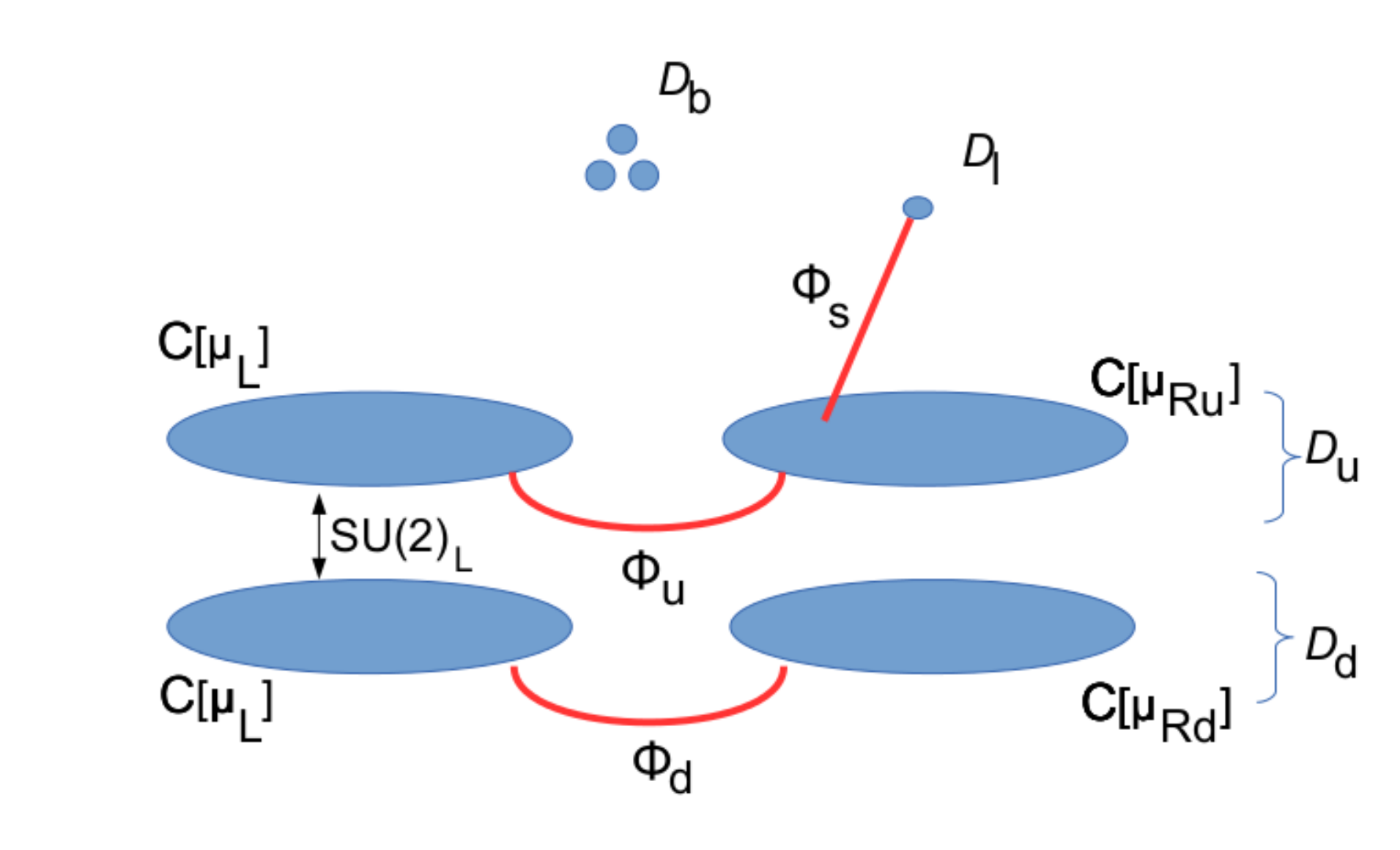}
 \end{center}
 \caption{Basic brane configuration for the standard model. }
 \label{fig:SM-branes}
\end{figure}

Now assume that some ``electroweak'' Higgs arise such that the 4 squashed  
$\cC[\mu_{i}]$ branes form two bound states 
\begin{align}
 \cD_u &=\cC[\mu_{L}]_u \cup_{\phi_u}  \cC[\mu_{Ru}] \nn\\
 \cD_d &= \cC[\mu_{L}]_d \cup_{\phi_d} \cC[\mu_{Rd}]  
 \label{Dud-branes}
\end{align}
as sketched in figure \ref{fig:SM-branes}.
For example, we could have $\cD_u = \cC[(1,0)] \cup  \cC[(0,1)]$ and 
$\cD_d = \cC[(1,0)] \cup \cC[(0,2)]$.
Dropping  indices, we can write the corresponding Higgs  suggestively as
\begin{align}
 \phi_d &\sim \sum \big(0|\mu_{L}\rangle_u + \varphi_{d} |\mu_{L}\rangle_d\big)
                   \langle\mu_{Rd}|_d  \cong \begin{pmatrix}
                                         0 \\ \varphi_{d} 
                                        \end{pmatrix}  \langle\mu_{Rd}|_d  \nn\\
 \phi_u &\sim \sum \big(\varphi_{u}|\mu_{L}\rangle_u  +  0|\mu_{L}\rangle_d\big)
                \langle\mu_{Ru}|_u \cong \begin{pmatrix}
                                         \varphi_{u} \\ 0
                                        \end{pmatrix} \langle\mu_{Ru}|_u
 \label{Higgs-doublets}
\end{align}
connecting some of their extremal weight states of $\cH_{L}$ and $\cH_{{R}}$
Thus $\varphi_{d}$ and $\varphi_{u}$ can be viewed as non-vanishing entries of two $SU(2)_L$ doublets
as in the MSSM  \eq{Higgs-doublets-2}, with 
 \begin{align}
 \tan\b = \frac{\varphi_u}{\varphi_d} \ .
\end{align}
We assume that the scale of the  $\cC[\mu]$ branes is 
much larger than the (electroweak) scale of the Higgs, $\varphi\ll r=1$
and  $\varphi\ll \varphi_S$,
so that we can neglect the back-reaction of the Higgs on the branes.
This defines the background under consideration.
The 3 coincident ``baryonic'' point branes $\cD_{b_j}, \ j = 1,2,3$ remain disconnected
from the rest. As discussed above, such squashed branes are solutions of our model.
The above Higgs are part of the zero mode sector, and  we simply assume  that they 
acquire some VEV.

Once these Higgs fields $\varphi_{d}$ and  $\varphi_{u}$ are switched on, 
the gauge symmetry is broken\footnote{Note that nonabelian VEVs in the scalar sector do reduce the 
rank of the gauge group.} to
$SU(3)_c \times U(1)_{Q} \times U(1)_{tr} \times U(1)_B$, where $B$ is the baryon number,
and 
\begin{align}
Q &:= \frac 12\big(\one_{{Ru}} +\one_{{Lu}}  -\one_{Rd} - \one_{Ld} + L - B \big) \ 
\label{Q-def}
\end{align}
is the electric charge generator.
Here $\one_{Lu}, \one_{Ld}$ indicate the $\cH_L$ which is part of $\cD_u$ and $\cD_d$, respectively.
Note that $Q$ is traceless provided $\dim\cH_{Ru} = \dim\cH_{Rd}$.
We will see that
$Q$ and $Y$  give the correct charge assignment of the standard model;
in particular, we note the  Gell-Mann-Nishjima formula
\begin{align}
 2Q-Y = \one_{Lu} - \one_{Ld} =: 2 T^3_L \ .
\end{align}
Thus the  low-energy broken gauge modes are given by three massive generators of
$SU(2)_L \times U(1)_Y$ identified as $W^\pm$ and $Z$, and the $U(1)_5$ mode generated by 
$T_5$.
To elaborate the masses of these low-energy gauge bosons, we decompose 
the Hilbert space of scalar fields on the two $\cC[\mu_L]$  as 
\begin{align}
 \End(\cH_L^2) = \End(\cH_{L}) \otimes \mmu(2)_L
 \label{End-H2-decomp}
\end{align}
where $\End(\cH_{L})$ are the functions on  $\cC[\mu_L]$.
Then the $W$ bosons arise from the $\msu(2)_L \subset \mmu(2)_L$-valued gauge fields 
which are proportional to $\one$ on $\cH_{L}$.
The  components of the $SU(2)_L \times U(1)_Y \subset U(N)$ gauge fields are accordingly given by  
\begin{align}
A_\mu(x) &= g_{N} (W_{\mu,i}(x)\, \tilde t_i  +  B_\mu(x) \tilde t_Y +  B_{\mu 5} \tilde t_5 ) \nn\\
 &= g \,W_{\mu,i}(x)\, t_i  + \frac 12 g' B_\mu(x) t_Y + g_5 B_{\mu 5} t_5, 
\end{align}
where
\begin{align}
  t_i &= c_L \tilde t_i =\one_{\cH_{L}} \otimes \frac 12\s_i \ , \qquad i = 1,2,3  \nn\\
  t_Y &= c_Y \tilde t_Y = \one_{Ru} -\one_{Rd} + L - B \ , \nn\\
  t_5 &= c_5 \tilde t_5 =  B-L+\Xi . 
\end{align}
They couple  to the fermionic zero modes
\begin{align}
 D_\mu \psi &= \big(\del_\mu -i [\cA_\mu,.]\big) \psi  
  = \big(\del_\mu -i g W_i t_i -i \frac{g'}2 B t_Y -i g_5 B_5\, t_5\big) \psi 
 \end{align}
and similarly to the Higgs fields $\phi_u, \phi_d$
\begin{align}
 D_\mu \phi &= \big(\del_\mu -i [\cA_\mu,.]\big) \phi  
  = \big(\del_\mu -i g W_i t_i -i \frac{g'}2 B t_Y -i g_5 B_5\, t_5\big) \phi \ .
\end{align}
As  explained in detail below, these reproduce precisely the 
couplings and charges of the standard model.
We can therefore identify 
 the gauge fields $W_i, B$, etc.  with those of the the standard model, where
$g$ is the  $SU(2)_L$ coupling constant, and $g'$ is the $U(1)_Y$ coupling constant. 
The coupling constants of the $SU(2)_L \times U(1)_Y$ gauge bosons are therefore given by
\begin{align}
 g &= \frac{g_N}{c_L}  , \qquad
 \frac 12 g' = \frac{g_N}{c_Y}, \qquad  g_5 = \frac{g_{\rm YM}}{c_5}  .
\end{align}
The appropriate normalization is obtained such that the Lagrangian of the gauge fields is
\begin{align}
  \cL_{\rm G} = - \frac 1{4} F_i^{\mu\nu} F_{\mu\nu}^i - \frac 1{4} B^{\mu\nu} B_{\mu\nu} + ...
 \label{gaugefields-SM}
\end{align}
i.e. $\tr\,\tilde t_i^2 = 1 =\tr\, \tilde t_Y^2$,
which gives
\begin{align}
 c_L  &= \sqrt{\frac 12\dim \cH_{L}} \ , \nn\\
 c_Y  &= \sqrt{\dim \cH_{Ru} + \dim \cH_{Rd} +\frac 43} \ ,\nn\\
 c_5  &= \sqrt{\dim \cH_{Ru} + \dim \cH_{Rd} + 2\dim \cH_{L} +\frac 43} \ .
\end{align}
Then the masses of the gauge bosons are obtained from
\begin{align}
 \cL_\phi[A] &= -\frac 12 \tr D_\mu \phi^\dagger D^\mu  \phi
 = -\frac 12 \tr [W_\mu,\phi]^\dagger[W^\mu,\phi] =: -\frac 12 W_\mu W^\mu  m^2_W
\end{align}
where the covariant derivatives of scalar fields \eq{Higgs-doublets} are explicitly
\begin{align}
 i D \phi_d &= [A,\phi_d] = (g W_a t_a + \frac 12 g' B + g_5 B_5) \phi_d  \nn\\   
   &=  \frac{\varphi_{d}}2 \begin{pmatrix}
         g(W_1+iW_2) \\ - g W_3 + g'B + 2g_5 B_5
        \end{pmatrix}  \langle\mu_R|_d 
     =  \frac{\varphi_d}2 \begin{pmatrix}
         g(W_1+iW_2) \\ - g_Z Z + 2g_5 B_5
        \end{pmatrix}  \langle\mu_R|_d ,  \nn\\
i D \phi_u &=  [A,\phi_u] = (g W_a t_a - \frac 12 g' B + g_5 B_5) \phi_u \nn\\
   &=  \frac{\varphi_u}2 \begin{pmatrix}
         g(W_1-iW_2) \\ g W_3 -g'B + 2g_5 B_5
        \end{pmatrix}  \langle\mu_R|_u
    =  \frac{\varphi_u}2 \begin{pmatrix}
         g(W_1-iW_2) \\  g_Z Z + 2g_5 B_5
        \end{pmatrix}  \langle\mu_R|_u \ .
 \label{covar-explicit}
\end{align}
The $Z$ boson is identified as the combination of $W_3$ and $B$ which acquires a mass, 
\begin{align}
 Z = g W_3 - g'B .
\end{align}
On the other hand, \eq{Q-def} guarantees that $U(1)_Q$ remains exactly massless, since
$\cD_{ul} = \cD_u\cup_{\phi_S}\cD_l$ and  $\cD_d$ are disconnected.
The masses are obtained from
\begin{align}
\cL_\phi[A] 
 &=  -  \frac{\varphi^2}8\big( g^2 (W_1^\mu W_{1\mu} + W_2^\mu W_{2\mu}) + (g^2+{g'}^2) Z^\mu Z_\mu  + 4 g_5^2 B_5^\mu B_{5\mu} \big) 
 \label{ew-boson-mass}
\end{align}
for $\varphi_u = \varphi \sin\b, \ \varphi_d = \varphi \cos\b$.
Here $\tr_N$ is evaluated using the explicit form \eq{covar-explicit} 
of $\phi$  connecting the 
extremal weight states of the squashed branes,
and does  {\em not} contribute any $N$-dependent factors\footnote{This is an essential improvement 
compared with the background in \cite{Steinacker:2014fja}, which lead to a factor $N$ at this point, 
and to an equal scale of the mirror fermions and $W$ bosons.}. 
We can then read off the tree-level $W$ and $Z$ bosons masses,
\begin{align}
 m_W^2 &= \frac 14 g^2 \varphi^2 , \qquad
 m_Z^2 = \frac 14 (g^2+{g'}^2)\varphi^2, \qquad m_5^2 = g_5^2 \varphi^2 \ .
 \label{W-mass}
\end{align}
All scales  are set by $m$.
Note that as long as $\varphi \ll r$,  $m_W^2$ is much lower than any of the higher KK gauge bosons
which start at $12 g_N^2 r^2$, where $12$ is the lowest eigenvalue of $\Box_X$ on $\cH_{(1,1)}$.
The photon and the $Z$-boson are now identified as usual 
\begin{align}
   \begin{pmatrix}
    Z\\  A
   \end{pmatrix} 
   &=  \begin{pmatrix}
                 \cos\theta_W & -\sin\theta_W \nn\\
                  \sin \theta_W& \cos\theta_W
                \end{pmatrix} 
                 \begin{pmatrix}
                 W_3 \\ B
                \end{pmatrix}  
= \frac{1}{\sqrt{g^2+{g'}^2}} \begin{pmatrix}
                 g & -{g'}\nn\\
                 g' & g
                 \end{pmatrix}
                \begin{pmatrix}
                 W_3 \\ B
                \end{pmatrix}  \ .
\end{align}
This gives the Weinberg angle
\begin{align}
 \sin^2\theta_W = \frac{{g'}^2}{{g'}^2+g^2} 
  = \frac{2\dim \cH_L}{2\dim \cH_L+\dim \cH_{Ru} + \dim \cH_{Rd} +\frac 43} \ .
\end{align}
E.g. for $\dim\cH_i = 3$ this gives $\sin^2\theta_W = 0.45$,
 for $\dim\cH_L=3, \ \dim \cH_R = 6$ this gives  $\sin^2\theta_W = 0.31$,
and for $\dim\cH_L=3, \ \dim \cH_R = 8$ this gives  $\sin^2\theta_W = 0.25$.
These are of course tree-level formulae which should be viewed as GUT values at very high energies.

These formulae have to be generalized in the 
presence of several Higgs components,
in particular $\tilde\varphi$ and $\varphi$ which couple to the mirror fermions
and standard-model fermions, respectively.
All of them contribute to the $W$ mass as above, and must be taken into account accordingly.

To put this into perspective,  consider briefly the coupling of the 
Higgs to the fermionic zero modes (which are discussed in detail below). We will see that
the off-diagonal fermionic zero modes connecting $\cD_l$ with $\cD_u$ or $\cD_d$ 
have the  structure 
\begin{align}
 \Psi &=  |s_i\rangle \psi_{12}\ , \qquad 
  \psi_{12} \sim  |\mu_{L,R}\rangle_{u,d} \langle 0|_{l}
\end{align}
and the Yukawa couplings among these arise from
\begin{align}
 \int d^4 x\, g_N \tr_N\bar\Psi\Gamma^a [\phi_a,.]\Psi 
  \ &\sim \ 2\int d^4 x\, g_N \varphi \bar\psi_{12} \psi_{12} \ 
\end{align}
The trace $\tr_N$ gives no extra factor since the fermions are made 
from coherent states, similar as the bosonic modes in \eq{ew-boson-mass}. Therefore
the fermion mass is given by
\begin{align}
 m_\psi \sim  g_N\varphi = c_L g \varphi \sim \sqrt{\frac 12\dim \cH_L}\, g\, \varphi \ ,
 \label{higgs-coupling-g}
\end{align}
which is much larger than the $W$ scale \eq{W-mass} for large branes with 
$\dim\cH \gg 1$. This implies that the mirror fermions can be much heavier than the 
$W$ scale, which is essential, and  resolves
one of the main issues in \cite{Steinacker:2014fja}. On the other hand, this also entails that 
the $SU(N)$  coupling $g_N$ is considerably larger than the 
electroweak coupling $g$. In the present paper, we 
focus on minimal or small branes.

In the next section we discuss the fermionic zero modes on such a background, and show
how the fermions of the standard model can arise.

\section{Fermionic zero modes}
\label{sec:fermions}

Now we turn to the fermionic zero modes, which provide the matter content of the 
low-energy field theory on the squashed $\cC_N[\mu]$  backgrounds. 
The basic results are obtained in \cite{Steinacker:2014lma,Steinacker:2014eua}, however 
we emphasize again their group-theoretical organization  which makes the 
relation with the scalar zero modes manifest.

The internal Dirac operator on a background  $X_\a$ describes a stack of  $\cC_N[\mu]$ branes
has the form
\begin{align}
\slashed{D}_{(\rm int)}^X \Psi &= \sum_{\b\in\cI} \Delta^\b [\Phi_\b,\Psi] 
  = \sqrt{2}\sum_{i=1}^3 \Big(\Delta^-_i [X_i^+,.] + \Delta_i^+ [X_i^-,.]\Big) 
 \label{Dirac-ladder}
\end{align}
where the spinorial ladder operators
\begin{align}
 2\Delta_1^+ =  \Delta_4 + i \Delta_5,   \quad
 2\Delta_2^+ =  \Delta_6 - i \Delta_7,   \quad
 2\Delta_3^+ =  \Delta_1 - i \Delta_2,   
\end{align}
and $\Delta_i^- = (\Delta_i^+)^\dagger$  satisfy
\begin{align}
\qquad  \{\Delta^-_i,\Delta_j^+\} = \d_{ij}. 
\end{align}
We recall the traceless generators $\t_i$ of the $U(1)_i \subset SU(3)_R$, 
and introduce their spinorial representation
\begin{align}
 \tilde\t_i &= \Sigma_{(i)} - \frac 12\sum_{j\neq i}\Sigma_j,  \nn\\
 \Sigma_{(i)} &= \frac 12[\Delta^-_i,\Delta_i^+]  = \frac 12\chi_i \ , \qquad i=1,2,3 \ 
 \label{sigma}
\end{align}
with $\tilde\t_1 + \tilde\t_2 + \tilde\t_3  = 0$. 
The 8 states in a  Dirac spinor of $SO(6)$ transform in the $(4)_L \oplus (\bar 4)_R$ of $SU(4)_R$, and
decompose into $(3)_L \oplus (\obar 3)_R \oplus (1)_L \oplus (1)_R$ under $SU(3)_R$. 
The 6 non-vanishing charges $\a=\pm\a_i$ of  $\tilde\t_i$ thus
form a regular hexagon $(3) \oplus (\obar 3)$ (just like the scalar fields $\phi_\a$)
\begin{align}
 |s_1,s_2,s_3\rangle  \equiv |\a\rangle, \qquad 
 2\tilde\t_i |\a\rangle  = (\a_i,\a)  |\a\rangle \ ,
 \label{spinor-a}
\end{align}
while two singlet  ``gaugino'' states 
\begin{align}
 \chi_L =|\uparrow\uparrow\uparrow\rangle, \qquad \chi_R =|\downarrow\downarrow\downarrow\rangle
 \label{gaugino-states}
\end{align}
have vanishing $\tilde\t_i$ charge  $\a=0$.
The spinors with $\a\neq 0$ have definite chirality determined by 
\begin{align}
 \chi |\pm\a_i\rangle = \chi_1 \chi_2 \chi_3 |\pm\a_i\rangle  
 = \pm |\pm\a_i\rangle \ = \,  \tilde\t |\pm\a_i\rangle
\end{align}
where $\tilde\t$ is the trace-$U(1)_R$ generator acting on the spinors corresponding to $\t$ \eq{tau-parity}.

\begin{figure}
\begin{center}
 \includegraphics[width=0.45\textwidth]{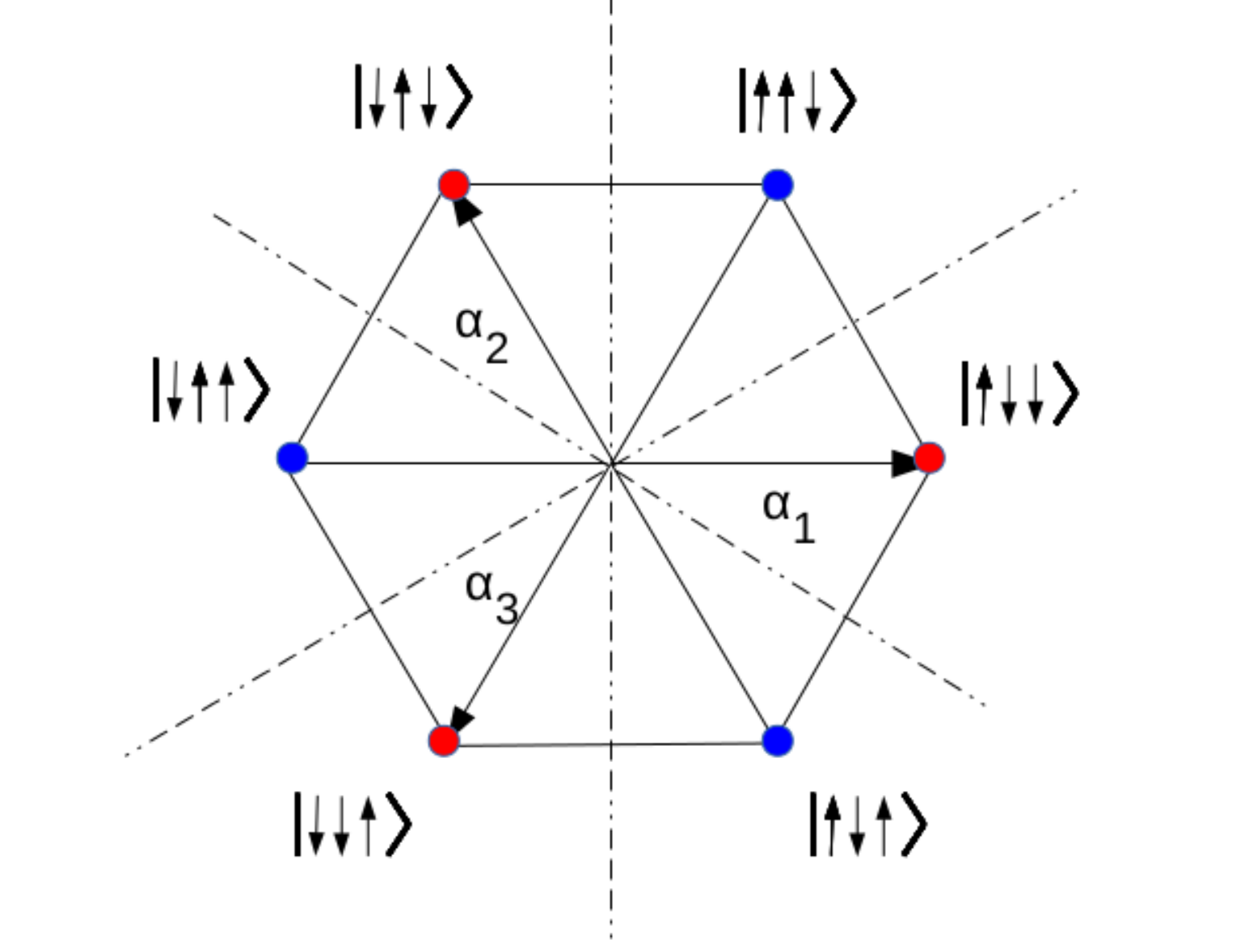}
 \end{center}
 \caption{Fermionic zero modes in weight space with root basis $\a_i$,  and chirality 
 sectors ${\color{red} C_L}$, ${\color{blue} C_R}$  indicated by color. The six Weyl chambers
 are separated by the dashed lines.}
 \label{fig:spin-states-roots-beta}
\end{figure}

Now we can exploit  the fact that the background preserves the $U(1)^K_i$ symmetries \eq{combined-U1-symm}.
This implies that the  Dirac operator $\slashed{D}_{(\rm int)}^X$ commutes with 
\begin{align}
\tilde K_i := 2\tilde\t_i - [H_{\a_i},.] ,  \qquad i=1,2,3 \  
\label{combined-U1-symm-fermions}
\end{align}
in analogy to \eq{combined-U1-symm}. 
As in section \ref{sec:scalar-zeromodes}, it follows that  for each irreducible 
$\cH_{\L} \subset \End(\cH)$, $\slashed{D}_{(\rm int)}^X$ has 6 zero modes labeled by $\a$
\begin{align}
 \slashed{D}_{(\rm int)}^X \Psi_{\a,\L'}^{(0)} &= 0,  \nn\\[1ex]    
 \Psi_{\a,\L'}^{(0)} &=|\a\rangle \, Y_{\l}\ , \qquad \L'=\a-\l, \qquad  Y_{\l} \in \cH_{\L} 
\end{align}
which are in one-to-one correspondence to the extremal weights $\L'$ of the $\tilde K_i$.
Here $Y_\l$ is some extremal weight vector in $\cH_{\L} \subset \End(\cH)$.
This follows from 1) the multiplicity of the extremal weight states is one, 
2) they are eigenvectors of $\chi$, and 3)
\begin{align}
 \slashed{D}_{(\rm int)}^X \, \chi &= - \chi \, \slashed{D}_{(\rm int)}^X \ .
\end{align}
In particular, these $\Psi_{\a,\L'}^{(0)}$ can be viewed as superpartners the 
bosonic regular zero modes $\phi_{\a,\L'}^{(0)}$ \eq{bosonic-zeromodes}, with the same charge $\a$
under $U(1)_i \subset SU(3)_R$.
For example, 
\begin{align}
 \Psi_{\a_3,\a_3-\L} ^{(0)} = |\downarrow\downarrow\uparrow\rangle \, Y_{\L} \ 
\end{align}
where $Y_{\L}$ is  the highest weight vector of $\cH_\L \subset \End(\cH)$. 
This can easily be verified directly using the 
form \eq{Dirac-ladder} of the Dirac operator,
together with
\begin{align}
  \Delta_1^- |\downarrow\downarrow\uparrow\rangle =  \Delta_2^- |\downarrow\downarrow\uparrow\rangle
   = \Delta_3^+ |\downarrow\downarrow\uparrow\rangle = 0 .
  \label{extremal-spin-states}
\end{align}
These  states are visualized\footnote{This picture differs from figure 1 in
\cite{Steinacker:2014eua} due to a different choice of roots.}  
in figure \ref{fig:spin-states-roots-beta}. 
They fall into chirality classes  $C_L$ and $C_R$ with
 well-defined internal chirality 
\begin{align} 
 \chi  \Psi_{\pm\a_i,\L'}^{(0)}  &= \tilde \t \, \Psi_{\pm\a_i,\L'}^{(0)}  \ =\ \pm \, \Psi_{\pm\a_i,\L'}^{(0)}  \ ,  \
 \label{chirality}
\end{align} 
 determined by the parity $\pm 1$ of the Weyl chamber of $\a=\pm\a_i$; recall that
the reflections along the $\t_i$ divide weight space into the 6 Weyl chambers of $\msu(3)_X$.
Now $\a$ is in the same Weyl chamber as $\L'=\a-\L$, 
and in the opposite Weyl chamber as the gauge charge $\L$
of the matrix wave-function $Y_{\L}$. 
Thus the chirality of $\Psi_{\a,\L'}^{(0)}$ is determined by the 
parity  of $\a$, hence\footnote{This is strictly true only for  $\L$ 
which are not on the border of two Weyl chambers. 
Otherwise, there are two zero modes with opposite chirality associated to $\L$.}  by $\L$. 
This is an important statement,
because it signals a chiral behavior of the low-energy gauge theory.
All this is consistent with the vanishing index of $\slashed{D}_{(\rm int)}^X$.

It turns out that there are no other fermionic zero modes besides 
 these extremal zero modes, except for the trivial gaugino modes $\chi_L, \chi_R$ with $\L=0$.
 The remaining fermionic modes 
(including the gaugino modes for  $\L\neq 0$)
 acquire ``Kaluza-Klein'' masses with scale set by $m$. 
In particular, there are no fermionic zero modes 
corresponding to the exceptional scalar zero modes, hence supersymmetry is manifestly broken 
even in the low-energy  spectrum.

So far, we only discussed the internal spinor structure of the zero modes. 
Taking into account the 10D  Majorana-Weyl condition 
$\Psi^C = \Psi = \Gamma\Psi$, this translates directly to the space-time spinor structure.
It is easy to see (cf. \cite{Steinacker:2014lma}) that the extremal 
modes $\Psi_{\a,\L'}$ and $\Psi_{-\a,-\L'}$ are related by
the internal charge conjugation and have opposite chirality,
\begin{align}
 C^{(6)} \Psi_{\a,\L'}^* = \Psi_{-\a,-\L'} .
\end{align}
Let us use the short notation $\Psi_i^\pm = \Psi_{\pm\a_i,\L'}^{(0)}$, where $\pm\a_i$ are the roots of $\msu(3)_X$.
Taking into account the  Majorana-Weyl condition, the 
corresponding solutions of the full Dirac operator have the form 
\begin{align}
 \Psi_i(x) = \Psi_i^+ \otimes \psi^i_+(x) + \Psi_i^- \otimes \psi^i_-(x),
 \label{MW-spinor-full}
\end{align}
where the four-dimensional spinors $\psi^i_\pm$ satisfy
\begin{align}
 \Di_{(4)} \psi_\pm^i(x) & = 0, &
 (\psi_\pm^i(x))^C & = \psi^i_\mp(x).
 \label{MW-explicit}
\end{align}
and have specific chirality
\begin{align}
 \gamma_5 \psi_\pm^i(x) & =  \pm \psi_\pm^i(x) \ .
 \label{chirality-4D}
\end{align}
This means that the $\psi_\pm^i$ are not independent,
as $\psi_+^i(x)$ determines $\psi_-^i(x)$.
We can  expand the general solution in terms of 
plane wave Weyl spinors $\psi^\pm_{i;k}(x)$ on $\R^4$ with  momentum $k$, 
\begin{align}
 \Psi_i(x) = \int \frac{d^3 k}{\omega_k} \big(\psi_{i;k}^+(x)  \Psi^i_+
                                      + \psi^{-}_{i;k}(x) \Psi^i_- \big) , \qquad i= 1,2,3 .
\end{align} 
This can be viewed in terms of three 4-dimensional Weyl spinors $\psi^{+}_i$, 
which naturally form 3 chiral supermultiplets with the corresponding bosonic zero modes.

Together with the relation between the internal chirality and the charges $\L'$ established above, it follows
that the fermionic zero modes cannot acquire any mass terms even at the quantum level, 
as long as the $U(1)_i^K$ is unbroken. 
There are simply no other modes available with 
the opposite $\L'$ and the same 4D chirality to form a mass term in 4 dimensions.
This holds even in the  presence of mass terms such as in \eq{V-soft}
or their fermionic analogs.

Since the above analysis is based entirely on group theory, the classification of zero modes
carries over immediately to stacks of branes.
The results can then be summarized  by stating that a quiver gauge theory arises on
stack of squashed branes $\oplus n_i\cC[\mu_i]$, with gauge group $U(n_i)$ on each node $\mu_i$
and arrows corresponding to chiral superfields $\Phi_{\a,\L'}$
labeled by the extremal weights $\L'$ obtained by adding the six non-vanishing weights
$\a\in \cW(1,1)$ to the (negative) weights of $\cH_\L \subset \Hom(\cH_{\mu_i},\cH_{\mu_j})$.
Fields with opposite weights are conjugates of each other.
The trivial modes $\L=0$ on each node lead to $\cN=4$ supermultiplets.
However, this quiver does not give the full story, as there are exceptional scalar zero modes, 
heavy fields, and non-supersymmetric interactions which arise from the parent theory.

We will  restrict ourselves to the fermionic zero modes 
$\Psi_{\a,\L'}^{(0)}$ from now on. 
We emphasize again that all fermionic zero modes come in 3 generations, except for the two 
gaugino modes $\chi_{L,R}$ which arise for $\L=0$.

\subsection{Higgs fields and Yukawa couplings on minimal branes}
\label{sec:higgs-Yukawas}

Adding a Higgs to the background $\Phi_\a = m(X_\a + \phi_\a)$,
the fermionic zero modes may acquire masses through Yukawa couplings arising from $\slashed{D}_{(\rm int)}^\phi$, 
\begin{align}
 m \Tr \obar\Psi \gamma_5\slashed{D}_{(\rm int)}^\phi \Psi
 &= m\Tr \obar\Psi \gamma_5 \Delta^\a[\phi_\a,\Psi] \ .
 \label{Yukawa}
\end{align}
These Yukawas are non-vanishing only if the $U(1)_i$ charges of the three fields under $K_i$ add up to zero,
which provides a strong constraint for these  couplings. Since the gaugino modes \eq{gaugino-states} 
arise only for the trivial $\L=0$ modes, they cannot contribute any non-vanishing Yukawas in the zero mode sector.
Together with the $U(3)_R$ symmetry, this implies 
that the non-vanishing Yukawas in the zero-mode sector have the following form 
\begin{align}
 \Tr \obar\Psi_{-\a_i}^{(0)} \gamma_5 \Delta^{\a_j}[\phi^{(0)}_{\a_j},\Psi_{\a_k}^{(0)}]\ \sim\ \varepsilon_{ijk} 
\label{tau-cons-yuk}
\end{align}
or its conjugate. In particular, {\em the $\tau$-parity of $\a_i,\a_j,\a_k$ are equal}.
However, we do not know which Higgs  assume a non-vanishing VEV. This should be determined largely by the 
cubic flux term \eq{cubic-flux}, while the quartic potential will stabilization the Higgs, 
as discussed in section \ref{sec:interactions}.
Note that the structure of the Yukawa coupling \eq{tau-cons-yuk} is very similar 
to the cubic  flux term, which also couples only modes with the same $\tau$-parity. 
Since the flux term is odd, it is plausible that 
non-trivial solutions with non-vanishing Yukawa couplings  arise,
with separate $\tau =  \pm 1$ sectors. The latter will correspond to the light sector and the mirror sector below.
However, a detailed analysis is beyond the scope of the present paper.
We will thus make some simplifying assumptions in the following, 
in an attempt to identify physically interesting configurations for such Higgs and Yukawa couplings.

Our first  assumption is that there are no Higgs modes {\em on} any given $\cC[\mu]$
(linking a brane with itself).
We restrict ourselves to
Higgs fields $\phi_\a$ arising  as links between 
branes $\cC[\mu_L]$ and $\cC[\mu_{R}]$ in \eq{Dud-branes}. 
This  suffices to exhibit the separation into light and mirror fermions.

\paragraph{Minimal branes and Higgs.}

We restrict ourselves to the minimal squashed $\C P^2$ branes in this paper,
with $\mu_L=(1,0)$ and $\mu_R=(0,1)$. 
Then among all possible Higgs modes linking $\cC[\mu_L]$ and $\cC[\mu_R]$ 
in \eq{zeromodes-bosons-LR},
we focus on the regular zero modes with $\L'\in \cW(1,2)$ 
\begin{align}
 \phi_{\a,\L'}^{(0)}\, &= \varphi_\a^{ij} |\mu^i_L\rangle\langle \mu_R^j| 
 \ \in  \  (0,1) \subset  (1,0) \otimes (1,0) , \quad \L'\in \cW(1,2)  
 \label{minimal-Higgs}
\end{align}
with antisymmetric\footnote{the symmetric combination is part of $(2,0)$.} 
$\varphi_\a^{ij} = - \varphi_\a^{ji}$,
and  the  $\L'\in\cW(2,1)$ modes 
\begin{align}
 \phi_{\a,\L'}^{(0)}\, &= \tilde\varphi_{\a}^{ij} |\mu_R^j\rangle\langle \mu^i_L| 
 \ \in  \  (1,0) \subset  (0,1) \otimes (0,1), \quad \L'\in \cW(2,1) 
 \label{minimal-Higgs-conj}
\end{align}
which are determined by conjugation.
They link adjacent weights $\mu^i_L$ and $\mu^j_R$ of $(1,0)$ and $(0,1)$,
interpreted as strings linking the
sheets of $\cC[\mu_L]$ and $\cC[\mu_R]$ \cite{Steinacker:2014lma}. 
We will ignore the remaining regular zero modes with $\L' \in \cW(3,1)$ 
and the 3 exceptional zero modes with $\L'\in\cW(2,0)$ and $\L\in\cW(0,1)$ here, since  
they would not lead to Yukawas between the fermionic zero modes 
relevant to the SM. However they may give a mass to some of the extra 
(unwanted) fermions which arise besides the standard-model fermions, and 
should be taken into account eventually in a more complete analysis.

\begin{figure}
\begin{center}
 \includegraphics[width=0.45\textwidth]{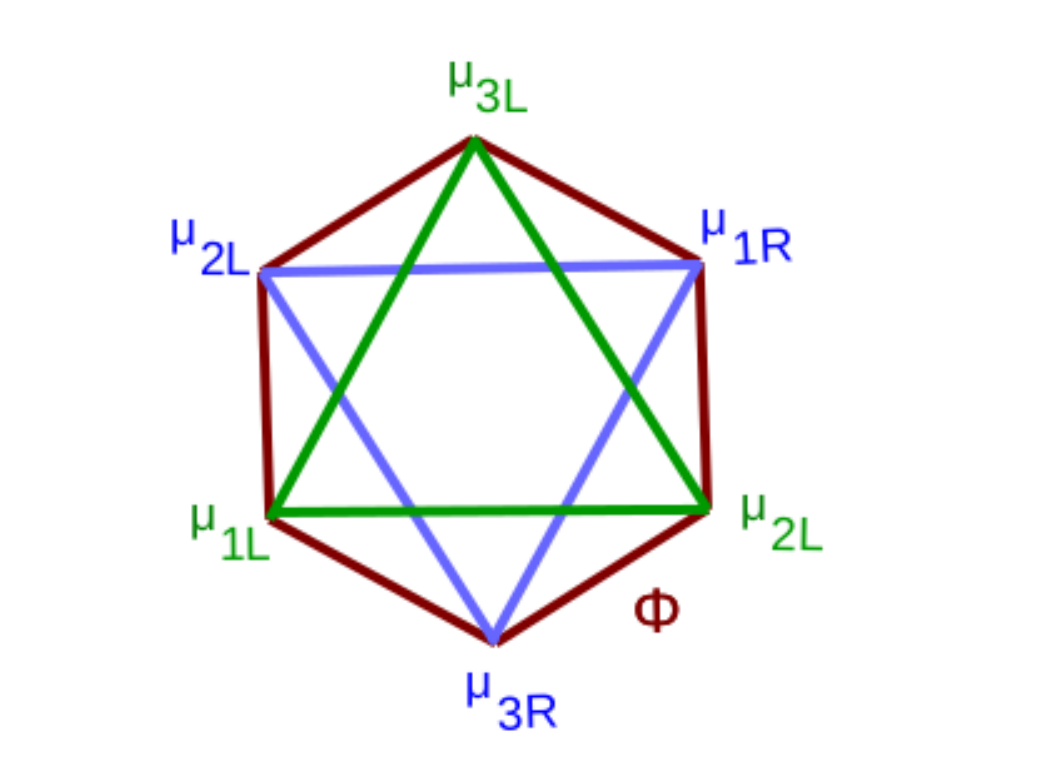}
 \end{center}
 \caption{${\color{green}\cC[(1,0)]} + {\color{blue}\cC[(0,1)]}$ with Higgs ${\color{red} \phi}$.}
 \label{fig:6-parameters}
\end{figure}

\begin{figure}
\begin{center}
 \includegraphics[width=0.5\textwidth]{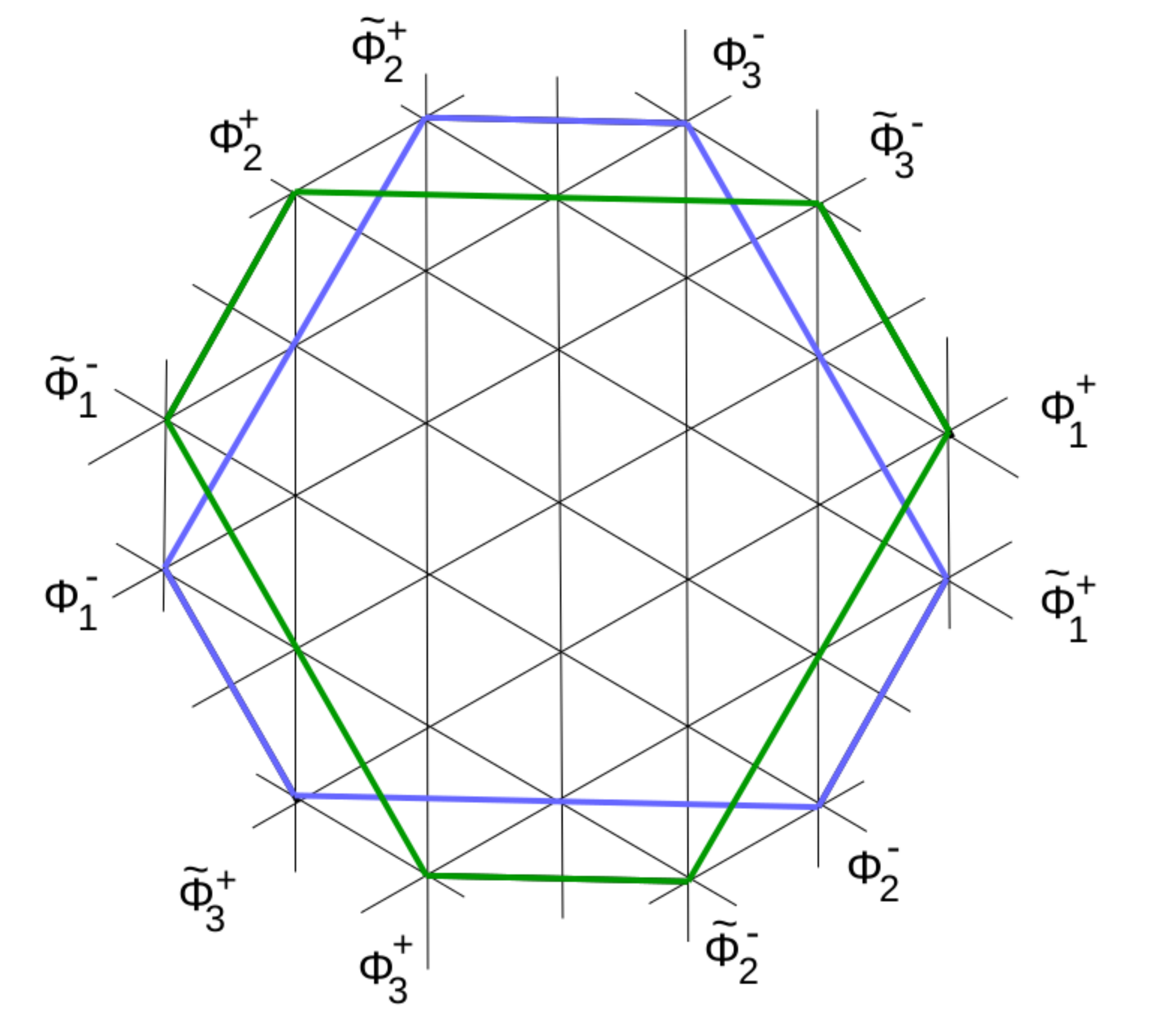}
 \end{center}
 \caption{Weight lattice of $\msu(3)$, with ${\color{blue}(1,2)}$ and 
 ${\color{green}(2,1)}$ irreps, and Higgs modes $\phi_{i}^\pm$ and 
  mirror Higgs $\tilde\phi_{i}^\pm$ in \eq{Higgs-i-structure}. 
 Their $\t$-parity is indicated by $\pm$.}
 \label{fig:weights-21}
\end{figure}

Consider these Higgs modes \eq{minimal-Higgs} in more detail. 
Since $(1,2)$ is the conjugate representation to $(2,1)$, 
the latter are determined by the 3+3 independent $(1,2)$ modes by conjugation. 
Equivalently, we can consider the three $\t=+1$ modes with $\L'\in \cW(1,2)$ 
and the three $\t=+1$ modes with $\L'\in \cW(2,1)$ as independent modes, 
which determine the remaining modes by conjugation.
Explicitly,
writing the $\cC[\mu_L] + \cC[\mu_R]$ background\footnote{The minus in the second term
reflects the fact that the generators of conjugate representations are related by minus transposition.} 
as 
\begin{align} 
 X_i^+ \equiv X_{\a_i} &=  r_i\big(|\mu^{i+1}_L\rangle \langle \mu^{i}_L| 
                                 - |\mu^{i}_R\rangle \langle \mu^{i+1}_R| \big), \qquad i=1,2,3 
\end{align}      
(cyclically)
where $\mu^i_R = - \mu^i_L$ and $\a_1 = \mu^2_L-\mu^1_L$ etc.,
these six independent Higgs fields are
\begin{align}    
  \phi_{i}^+  \, \equiv \,  \phi_{+\a_i,\L'}^{(0)} &= \varphi_{i}
(|\mu^{i-1}_R\rangle \langle \mu^{i+1}_L| - |\mu^{i+1}_R\rangle \langle \mu^{i-1}_L|) , \qquad \L'\in \cW(1,2)  \nn\\
  \tilde\phi_{i}^+  \, \equiv \,  \tilde\phi_{+\a_i,\L'}^{(0)} &= \tilde\varphi_{i}\, 
    (|\mu^{i-1}_L\rangle \langle \mu^{i}_R| - |\mu^{i}_L\rangle \langle \mu^{i-1}_R|) , \qquad \L'\in \cW(2,1)
\label{Higgs-i-structure}
\end{align} 
for $i=1,2,3$, which determine their conjugates $\phi_{i}^-$,  $\tilde\phi_{i}^-$
(see figure \ref{fig:6-parameters} and \ref{fig:weights-21}). 
The superscripts $\pm$ indicate the $\t$-parity $\t = \pm 1$.
Hence 
they are parametrized by 3 + 3 (complex) fields $\varphi_i$ and $\tilde\varphi_i$,
which will be referred to as ``Higgs'' and ``mirror Higgs''.

Now {\em  we assume that only  $\tilde\varphi_{i}\neq 0$, or more generally
$|\tilde\varphi_{i}| \gg |\varphi_{i}|$}.
In other words, the $\t$-parity  of the Higgs with $\L'\in\cW(2,1)$  is positive,
and the $\t$-parity of the  Higgs with $\L'\in\cW(1,2)$ is negative.
This is the crucial assumption, which will lead to a chiral low-energy theory.
It is reasonable, because the
flux term only couples fields with the same $\t$-parity;
however a detailed investigation is left for future work.
We will see that under this assumption, the ``mirror Higgs'' $\tilde\phi_{i}$ 
gives a large mass to the ``mirror'' sector of the standard model, 
leaving the chiral standard model with massless chiral fermions (and some extra fields) at low energies.
The (small) $\phi_{i}$ modes then
play the role of the low-energy Higgs, giving mass to these standard-model fermions as usual.

Finally, consider the $\phi_S$ Higgs \eq{s-Higgs} connecting a minimal $\cC[\mu_R]$ brane
with a ``point'' brane $\cD_l=\cC[0]$. 
Similar as the Higgs connecting minimal branes, we can organize them in terms of 3+3 modes with $\t$-parity $\t=+1$
\begin{align}
 \phi^{+}_{iS} &\equiv \,  \phi_{\a_i S}^{(0)} = \varphi_{i S}|\mu^{i+1}_R\rangle \langle 0|, \qquad \L'\in\cW(1,2) \nn\\
 \tilde\phi^{+}_{iS} &\equiv \,  \tilde\phi_{\a_i S}^{(0)} = \tilde\varphi_{i S}|0\rangle \langle \mu^{i}_R|, 
   \qquad \L'\in\cW(2,1)
\label{Higgs-S-structure}
\end{align}
in the basis \eq{Higgs-i-structure}, and their conjugates. 
Those may be switched on independent of each other.
There are also 3 exceptional scalar zero modes with $\L' \in \cW (0,2)$, which we will ignore here.

\paragraph{Fermions between branes and points.}

Now consider the fermionic zero modes in more detail.
The  zero modes   linking  $\cC[\mu_L]$ and $\cC[\mu_R]$  with a point brane $\cC[0]$ 
are given by
\begin{align}
\Psi_{\a,\L'}^{(0)} &= |\a\rangle\psi_{\mu^\a_L 0}, \qquad 
 \psi_{\mu^\a_L 0}\,= |\mu^\a_L\rangle\langle 0|  \ \in  \  (1,0)\ , \quad \L'=\a-\mu^\a_L \in \cW(2,1)   \nn\\
 \tilde\Psi_{\a,\L'}^{(0)} &= |\a\rangle \psi_{\mu^\a_R 0}, \qquad 
 \psi_{\mu^\a_R 0}\,= |\mu^\a_R\rangle\langle 0|  \ \in  \  (0,1)\ , \quad \L'=\a-\mu^\a_R\in \cW(1,2) \ .
 \label{zeromodes-fermions-L0-1}
\end{align}
These are 6 zero modes with $\L' \in \cW (2,1)$ and 6  zero modes with $\L' \in \cW (1,2)$,
with chirality determined by the $\t$-parity of $\a$. 
The Yukawa coupling $\Tr \obar\Psi_\b^{(0)} \gamma_5 \Delta^\a[\phi_\a^{(0)},\Psi_\g^{(0)}]$
of  two such fermionic zero modes 
with the Higgs fields $\phi_{\a}^{(0)}$ is
non-vanishing only if the $U(1)^K_i$ charges $\L'$ of $\phi_\a$ and $\Psi_\g$ add up to that of 
$\Psi_\b$. A direct inspection of the $\msu(3)$ weight lattice
(see figure \ref{fig:weights-21})
shows that $\cW(2,1) + \cW(2,1) = \cW(1,2)$ has indeed solutions provided
the parities of $\g$ and $\a$ are equal and opposite to that of $\b$.
There are no other couplings among these modes, consistent with 
the general discussion following \eq{tau-cons-yuk}.
Together with the above assumption on the Higgs, this means that Yukawa couplings arise only between 
left-handed $\Psi_{+\a_i,\L'}^{(0)}$  with $\L'\in\cW(2,1)$ 
and right-handed $\tilde\Psi_{-\a_j,\L'}^{(0)}$ with  $\L'\in\cW(1,2)$,
mediated by $\tilde\phi_{+\a_k}$ with $\L'\in\cW(2,1)$.
Hence the fermions \eq{zeromodes-fermions-L0-1} separate into ``mirror'' fermions 
\begin{align}
 \{\psi^{\rm mirror}\} = \{\Psi_{+\a_i}^{(0)}; \  \chi = +1  \}\ \cup \ 
  \{\tilde\Psi_{-\a_i}^{(0)}; \  \chi = -1  \}
\end{align}
which acquire a large mass of order $\tilde\varphi$, while
the remaining  modes remain massless and constitute the 
``light'' sector 
\begin{align}
 \{\psi^{\rm light} \} = \{\Psi_{-\a_i}^{(0)}; \  \chi = - 1  \}\ \cup \ 
  \{\tilde\Psi_{+\a_i}^{(0)}; \  \chi = +1  \} \ .
\end{align}
To summarize, the light fermions are left-handed links from $\cC[0]$ to $\cC[\mu_L]$,
and right-handed links from $\cC[0]$ to $\cC[\mu_R]$.
If $\varphi_i\neq 0$, then these  light fermions couple among themselves and
 acquire a mass of order $\varphi$.

Now the crucial point is that the zero modes \eq{zeromodes-fermions-L0-1} with $\L'\in\cW(2,1)$ 
and $\L'\in\cW(1,2)$ are distinguished by their gauge charges: 
\begin{align}
\varXi  \begin{pmatrix}
                            \Psi_{\a}^{(0)}  \\
                            \tilde\Psi_{\a}^{(0)}
                           \end{pmatrix}
 = \begin{pmatrix}
    1 & 0 \\
    0 & -1
   \end{pmatrix}
   \begin{pmatrix}
                            \Psi_{\a}^{(0)}\\
                            \tilde\Psi_{\a}^{(0)}
                           \end{pmatrix}, \qquad
   \varXi  =  \one_L - \one_{R}
\end{align}
where $\varXi \in \msu(N)$ is the gauge generator (spontaneously broken by $\tilde\phi$)  which assigns the charges $\pm 1$ 
to the branes  $\cC[\mu_L]$ and $\cC[\mu_R]$.
Combining these results, we conclude that 
\begin{align}
 \chi \psi^{\rm mirror} &=  \varXi \psi^{\rm mirror}   \nn\\
 \chi \psi^{\rm light}  &= - \varXi \psi^{\rm light} 
 \label{chi-Xi-zeromodes}
\end{align}
hence 
\begin{align}
\g_5 \psi^{\rm light} = -\varXi  \psi^{\rm light} 
\label{chiral-light-4D}
\end{align}
using \eq{chirality-4D}.
This means that the low-energy fermions $\psi^{\rm light}$ are chiral as seen by the 
spontaneously broken gauge fields, just like the fermions in the standard model (in the broken phase).
The basic result \eq{chi-Xi-zeromodes} will be verified numerically in section \ref{sec:numerics}, and the
relation with the standard model will be made more specific below.

Finally assume that in addition $\phi^S$ \eq{Higgs-S-structure} is switched on, connecting  
$\cC[\mu_R]$ with $\cC[0]$. This will 
induce Yukawa couplings of $\psi_{\mu_R 0}$ with fermions on $\cC[\mu_R]$
and on $\cC[0]$, 
and possibly Yukawa couplings of $\psi_{\mu_L 0}$ with  $\psi_{\mu_L \mu_R}$.
Switching on $\phi^S\neq 0$ or $\tilde\phi^S = 0$ selectively, 
this should give a mass to $\psi_{\mu_R 0}$ while leaving the light fermions  $\psi_{\mu_L 0}^{\rm light}$ 
 massless. This is desirable since it will give mass to $\nu_R$, however a detailed investigation is left for 
 future work.

\paragraph{Fermions on and between branes.}

Now consider the fermionic zero modes linking different  $\cC[\mu]$ branes.
They are in one-to-one correspondence with the 
regular scalar zero modes discussed above.
In particular, the zero modes connecting 
two minimal branes $\cC[\mu_L]$ with $\cC[\mu_R]$ have the form
\begin{align}
 \Psi_{\a,\L'}^{(0)} =|\a\rangle\psi_{\a}, \qquad 
 \psi_{\a}\, = \psi_{\a}^{ij}|\mu_L^i\rangle\langle \mu_R^j| 
 \ \in  \ (1,0) \otimes (1,0) = (2,0) + (0,1) 
 \label{zeromodes-fermions-LR-minimal}
\end{align}
corresponding to \eq{zeromodes-bosons-LR}, 
where $|\a\rangle$ stands for the spinor \eq{spinor-a} with weight $\a$.
This leads to 6  zero modes with $\L' \in \cW (3,1)$ 
and 6  zero modes with $\L' \in \cW (1,2)$. 
The latter are the superpartners of the  Higgs fields 
\eq{minimal-Higgs-conj}.

There are also fermionic zero modes on some minimal branes $\cC[\mu]$, 
\begin{align}
 \Psi_{\a,\L'}^{(0)} = |\a\rangle \psi_{\a} ,
 \qquad \psi_{\a} =\psi_{\a}^{ij} |\mu^i\rangle \langle\mu^j| \ .
\end{align}
Six of these have $\L'\in\cW(2,2)$, and 8 are trivial modes $\psi_{\a} =\one$ 
with $\L'\in\cW(1,1)$.
If $\cC[\mu_L]$ and $\cC[\mu_R]$ are connected with a Higgs $\phi_{\a}^{(0)}$ as above,
then Yukawa couplings with structure
$\tr\obar\psi_{\mu_L\mu_R}\phi_{\a}^{(0)}\psi_{\mu_R\mu_R}$ 
and $\tr\obar\psi_{\mu_R\mu_L}\phi_{\a}^{(0)}\psi_{\mu_L \mu_L}$  arise,
giving mass to some of these fermions.
Rather than attempting a detailed analytical explanation here, we 
will analyze this numerically in the next section.

\section{Standard model fermions from  branes}

Now we apply these results to the brane configuration for the standard model \eq{Dud-branes}.
Consider the off-diagonal fermions
linking the $2\times\cC[\mu_L] +\cC[\mu_{Rd}] +\cC[\mu_{Ru}] + \cD_l +3\times \cD_b$ branes.
In the basis $(Lu,Ld,Rd,Ru,l,b_i)$, we denote these fermions as 
\begin{align}
 \Psi =
 \begin{pmatrix}
  *_{2} & \widetilde H_d  & \widetilde H_u  &  l_L & Q_L \\
   &  *  & e' & e_R & d_R \\
   &    & * & \nu_R & u_R  \\
   & & ~~~ & * &  u' \\
&  & & & *_3
\end{pmatrix} .
\label{fermion-matrix}
\end{align}
The fermions of the SM arise as links between the point branes  $\cD_l$ and $\cD_b$
and $\cD_u =\cC[\mu_{L}]_u \cup \cC[\mu_{Ru}]$ resp.
$\cD_d = \cC[\mu_{L}]_d \cup \cC[\mu_{Rd}]$, i.e. 
\begin{align}
Q_L = \begin{pmatrix} u_L \\ d_L   \end{pmatrix}, \qquad
l_L = \begin{pmatrix} \nu_L \\ e_L   \end{pmatrix},  
\end{align} 
as well as the right-handed leptons and quarks.
Furthermore there are slots for the Higgsinos $\widetilde H_u, \ \widetilde H_d$ as in the MSSM.
The charge generators 
\begin{align}
 Q &= \frac 12\diag(1,-1,-1,1,1,-\frac 13), \nn\\
  Y &= \diag(0,0,-1,1,1, - \frac 13) 
\end{align}
assign the following
quantum numbers $(Q,Y)$ to these off-diagonal modes 
\begin{align}
 (Q,Y)|_\Psi =
 \begin{pmatrix}
  * \ \ &  \begin{pmatrix} (1,1) \\ (0,1) \end{pmatrix}  &  \begin{pmatrix} (0,-1) \\ (-1,-1) \end{pmatrix}  & \begin{pmatrix} (0,-1) \\ (-1,-1) \end{pmatrix}  & \begin{pmatrix} (\frac 23,\frac 13) \\ (-\frac 13,\frac 13)\end{pmatrix}  \\[2ex]
   &  *  & (-1,-2)   &(-1,-2) & (-\frac 13,-\frac 23) \\[1ex]
   &    & * & (0,0) & (\frac 23,\frac 43)  \\[1ex]
   & & ~~~ & * & (\frac 23,\frac 43) \\
&  & & & *
\end{pmatrix} 
\label{charges}
\end{align}
(the $SU(3)$ assignment is obvious, hence dropped).
All quantum numbers of the standard model are correctly reproduced (cf. \cite{Grosse:2010zq,Chatzistavrakidis:2011gs}), 
and 3 families arise automatically 
due to the $\Z_3$ symmetry.
The Yukawa couplings  may of course break the 
$\Z_3$, and will be discussed below.
Thus the leptons arise as fermions linking $\cD_u$ or $\cD_d$ with $\cD_l$,
and the quarks arise as  fermions linking $\cD_u$ or $\cD_d$ with $\cD_l$.

All these modes have scalar superpartners given by the regular scalar zero modes. In particular,
the two Higgs doublets\footnote{Unfortunately there is a conflict 
with the standard  particle physics conventions, where the role of the $H_{u,d}$
is reversed, as is seen from their quantum numbers \eq{charges}. 
The present notation is  forced upon us by \eq{Dud-branes}.}
\begin{align}
 H_d = \begin{pmatrix}
         0 \\ \varphi_d
        \end{pmatrix}, \qquad 
 H_u = \begin{pmatrix}
         \varphi_u \\ 0
        \end{pmatrix}
  \label{Higgs-doublets-2}
\end{align}
with  $Y(H_d) = 1$ (as in the standard model) and
$Y(H_u) = -1$ (as in the MSSM) fit into the above matrix structure as 
\begin{align}
 \phi^a = \begin{pmatrix}
         0_2 & H_d & H_u & 0 & 0 \\
         & 0 & 0 & 0 & 0\\
         && 0 & \varphi_S & 0\\
         &&& 0 & 0 \\
         &&&&0
        \end{pmatrix}
 \ =\ \begin{pmatrix}
          0 & 0 & 0 & \varphi_u & 0&0\\
          0&0 & \varphi_d & 0 &  0&0\\
           0 & \varphi_d^{\dagger} & 0& 0 &0&0 \\
          {\varphi_u}^{\dagger} & 0 & 0& 0  & \varphi_S & 0\\
          0&0&0&  \varphi_S^\dagger & 0 & 0 \\
          0&0&0&0 & 0 & 0\end{pmatrix} .
\label{higgs-matrix}
\end{align}
This  indeed leads to the desired pattern of electroweak symmetry breaking, as shown in section \ref{SM-branes}.
We  also exhibit the ``sterile'' Higgs $\varphi_S$, which is a  singlet under the 
standard model gauge group, occupying the same slot as $\nu_R$.
The chiralities and masses of the fermions depend on the Higgs expectation values.
We will see in the next section that for  $\tilde\varphi \gg \varphi$,
the low-energy fermions linking point branes with $\cC[\mu_L]$ are left-handed,
and those linking with $\cC[\mu_R]$ are right-handed.
The fermions with the opposite chiralities 
-- which necessarily exist due to the vanishing index in $\cN=4$ SYM  --
acquire a large mass terms of order $\tilde\varphi$, and are  therefore invisible at low energies.
Thus the fermions of the standard model have indeed  the appropriate chirality at low energies, 
as suggested by their names $l_L,e_R$ etc. 
Finally, recall that 
the modes in the lower-diagonal part of the matrices 
are identified by the MW condition with the upper-diagonal ones, and therefore do not constitute 
independent degrees of freedom.

It is remarkable that no exotic charges arise:
all the charges in \eq{charges} 
correspond to the charges of the standard model, extended by the second Higgs doublet
and the sterile $\nu_R$. Thus we recover all fermions in the MSSM (including e.g. gluinos, winos and binos),
extended by
\begin{align}
 u' \sim |0\rangle_l\langle 0|_b
 \label{uprime}
\end{align}
which has the same quantum numbers as the $u_R$ quarks
(but it comes with both chiralities), and
\begin{align}
 e' \sim |\mu_{Rd}\rangle \langle\mu_{Ru}| 
\end{align}
which has the same quantum numbers as  $e_R$. 
This degeneracy
can be understood by viewing  $\cD_{\mu_{Ru}} \cup \cD_l$ as a single brane 
linked via $\phi_S$.
Thus $u'$ may mix with $u_R$, and $e'$ with $e_R$, 
and similarly $\nu_L$ may mix with neutral Higgsino $\tilde \varphi_u$ at low energies.
The $e'$ can be viewed as superpartner of the would-be $SU(2)_R$ gauge bosons connecting
$\cC[\mu_{Ru}]$ and $\cC[\mu_{Rd}]$ if $\mu_{Ru} = \mu_{Rd}$.
Finally there are a number of fermions which are neutral under the SM gauge groups. 
This includes  the superpartner of the (broken) $U(1)_L$ gauge field 
\begin{align}
  \l :=|0\rangle_l\langle 0|_l \ ,
 \label{lambda}
\end{align}
some diagonal ``neutralino'' modes on $\cD_u$ and $\cD_d$, and of course $\nu_R$.
The multiplets come in several incarnations corresponding to different $\L'$ modes, 
which may acquire a mass from the Higgs(es). 
This is discussed next.

One might absorb the extra fields $e'$ and $u'$ by replacing 
$\cC[\mu_{Ru}] \cup_{\phi_S} \cD_l$ by a single brane 
$\cC[\tilde\mu_{Ru}]$, as discussed in section \ref{SM-branes}. 
However, then there are typically extra modes connecting the extended branes.
The main reason for keeping $\cD_l$
separate is to  keep things transparent by
working with minimal branes $\cC[\mu_i]$  
and point branes $\cD_{l,b}$.

\subsection{Chiral fermions and Yukawas on the standard model branes}
\label{fermions-yukawas}

Now we apply our results on the Yukawa couplings 
to this brane configuration,
with $\mu_L = (1,0)$ and $\mu_R = (0,1)$. 
In particular, we assume that $\cD_u = \cC[\mu_{L}]_u \cup \cC[\mu_{Ru}]$ 
are linked by Higgs $\phi_\a, \tilde\phi_\a$ as above, 
and similarly for $\cD_d = \cC[\mu_{L}]_d \cup \cC[\mu_{Rd}]$ .


Consider first the fermions linking the point branes $\cD_l, \cD_b$ with $\cD_u$ or $\cD_d$.
Assuming that $\tilde\varphi \gg \varphi$, 
the results of the previous section imply that 
these separate into light fermions with masses of order $\varphi$,
and heavy mirror fermions with masses of order $\tilde\varphi$.
This leaves only the light fermions at low energy, which comprise 
left-handed fermions linking $\cC[0]$ to $\cC[\mu_L]$,
and right-handed fermions linking $\cC[0]$ to $\cC[\mu_R]$.
They correspond to  the standard-model-like chiral leptons and quarks.
The mirror fermions have the same S.M. quantum numbers but the opposite chiralities,
distinguished by the $U(1)_{K_i}$ quantum nmbers.
Due to the simple mode decomposition\footnote{This holds also for genral non-minimal branes,
as long as $\cD_l, \cD_b$ are point branes.}
\eq{zeromodes-fermions-L0-1}, we get precisely 
the same quark and lepton with their superpartners as in the MSSM,
plus their mirror modes at higher energies (which also form supermultiplets).

Now consider the low-energy fermions which arise  on the $\cD_d$ and $\cD_u$ branes 
(i.e. in the upper-left
$4\times 4$ block in \eq{fermion-matrix}). This includes 
the superpartners of the electroweak sector, such as Higgsinos, Winos, Binos, charginos
and neutralinos, as well as the $e'$. They come in different multiplets corresponding
to the different $\L'$ modes in \eq{zeromodes-fermions-LR-minimal}.
Their precise Yukawa couplings and masses
in this sector are rather complicated and will not 
be discussed in detail here; some illustrative numerical results are given in the next section.
Since the  $\L=0$ modes come as $\cN=4$ multiplets, there are also 3 generations of
chiral supermultiplets corresponding to the $W$ and $Z$ bosons. 
The numerical results indicate that some but 
not all of these acquire a mass from the mirror Higgs $\tilde\varphi$,
which suggests that some of the other Higgs discussed in section \ref{sec:higgs-Yukawas}
should also acquire a VEV. We leave this for further investigations.

Finally some fermionic would-be zero modes
arise within the 4 point branes $\cD_l + 3\cD_b$. This includes
gluinos with $Y=Q=0$, 
the color triplet $u'$ \eq{uprime} which is similar to $u_R$, and the singlets 
$\l$  on $\cD_l$ \eq{lambda}. 
We only discuss some aspects here, postponing a detailed analysis to future work.
First, the Higgs $\phi_S$ with $\L' \in\cW(2,1)$  should  lead to a Yukawa coupling of the 
$\nu_R$ with the $\l$ modes with $\L'\in\cW(1,1)$, 
and give a large mass to both $\nu_R$ and $\l$
(except for the two gaugino polarizations \eq{gaugino-states} of $\l$).
Similarly, the $u'$ might couple to $u_R$ via $\phi_S$,
(except for the two gaugino polarizations of $u'$), giving a mass to  $u'$ and $u_R$.
 It is tempting to speculate that the 
large Yukawa couplings of the top quark may be related to the 
presence of $\phi_S$.
The fate of the two gaugino polarizations of $u'$ and $\l$ is unclear. In any case,
the sector containing $\cD_l, \cD_b$ and  $\cC[\mu_{uR}]$  is rather complex and
should be studied elsewhere.

Due to the different parity modes of $\phi_S$ and $\tilde\phi_S$, it is possible that 
e.g.  $\nu_R\, $ acquires a large mass but not its mirror $\tilde \nu_R$.
Then the seesaw mechanism would apply to the physical neutrinos but not to the mirror
neutrinos, and no new massless neutrinos would be introduced.

The main result here is the separation of leptons and quarks into light chiral and heavy mirror sectors,
assuming a suitable Higgs configuration.
The crucial decoupling of the light and mirror sector is  guaranteed 
by the global $U(1)_i^K$ symmetry, and persists in the presence of 
explicit mass terms respecting that symmetry, 
such as in the $\cN=1^*$ model discussed in appendix \ref{sec:N=1}.
This mechanism will be verified numerically below, along with some illustrative sample computations for the 
remaining sectors.

\paragraph{Extra $U(1)$'s and anomalies.}

In the presence of chiral fermions, 
the  $U(1)_i$ gauge fields arising on backgrounds consisting of several $\cC[\mu_i]$ branes
deserve special attention. Consider first the  field theory setting at hand. 
Assuming  that the mirror Higgs $\tilde \varphi$ is much larger than the light Higgs $\varphi$ as above,
some of these $U(1)$'s aquire anomalous contributions from the low-energy sector
(i.e. after integrating out the massive mirror fermions).
In the present brane configuration this is the case for the $U(1)_B$ and the $U(1)_5$ fields. 
However, this anomaly from the light sector is canceled precisely by the anomaly from the mirror sector, 
so that there is no overall anomaly, in accordance with t'Hooft 
anomaly matching and the fact that $\cN=4$ SYM has no gauge anomalies.
In particular, the $U(1)_B$ gauge field remains massless in this model.


%

To get a better perspective on these $U(1)$'s, 
it is useful to recall the analogous situation in string theory. The present 
softly broken $\cN=4$ SYM model would arise ``locally" e.g. on stacks of $N$ $D3$-branes in a suitable flux background, 
cf. \cite{Camara:2003ku,Grana:2003ek,Aldazabal:2000cn}. Then some of these $U(1)$'s are anomalous at low energy 
and  acquire a mass through a St\"uckelberg mechanism 
\cite{Coriano':2005js,Coriano:2006xh,Coriano:2007fw,Morelli:2009ev,Kors:2005uz,Anastasopoulos:2006cz}, absorbing the 
corresponding St\"uckelberg field or axion which arises from the RR fields in string theory. 
This is a manifestation of the Green-Schwarz anomaly cancellation. 
In the present $\cN=4$ field theory setting, there is no manifest axion,
hence the correspondence with the string theory case is not fully realized.
However for analogous backgrounds in noncommutative $\cN=4$ in the IKKT matrix model, 
axion-like fields do arise \cite{Steinacker:2007dq}, as expected from the relation with string theory.
One may then hope that an analogous St\"uckelberg mechanism applies  and renders some of these $U(1)$'s massive,
however the details remain to be understood.



\subsection{Aspects of the Higgs potential}
\label{sec:interactions}

Now consider the interacting potential for the Higgs i.e. the scalar zero modes $\phi^{(0)}$ 
on a background solution $X$. The linear term in $\phi$ vanishes, 
so that the effective potential for $\phi$ obtained from \eq{V-full-expand} is 
\begin{align}
 V(\phi)
 &= \tr \big( \frac 12 \phi^\a  \big(\Box_X + 2 \Di_{\rm diag} \big) \phi_\a 
 + (X^\a + \frac 14\phi^\a)\Box_\phi \phi_\a  - \frac 12f^2 \big) + V_{\rm soft}(\phi) .
\end{align}
The cubic interaction arising from the quartic term can be written in different ways
\begin{align} 
  \tr X^\a\Box_\phi \phi_\a 
 &=  \tr[X_\a,\phi_\b][\phi^\a,\phi^\b] \nn\\
  &=  -\tr \phi_\b [[\phi^\a,\phi^\b],X_\a] 
  =  \tr \phi_\b \Big([[\phi^\b,X_\a],\phi^\a]  + [[X_\a,\phi^\a],\phi^\b] \Big)\nn\\ 
  &=  -\tr \phi_\b [[\phi_{-\b},X_\a],\phi^\a] 
  \label{V-phi-rewrite}
\end{align} 
using the Jacobi identity,  $\phi^\b = \phi_{-\b}$, and
the gauge condition $f = [X_\a,\phi^\a] =0$. The latter is a special case of the following identities
\begin{align}
 [X_\a,\phi_\b^{(0)}]=0\qquad \mbox{if}\quad  \a+\b \in \cI \quad \mbox{or}\ \  \a+\b=0 
 \end{align}
 and
 \begin{align}
 [X_{\a},\phi_{-\b}^{(0)}]=0\qquad \mbox{if}\quad  \a-\b \in \cI \quad \mbox{or}\ \  \a-\b=0
 \label{extremal-char}
\end{align}
for the regular zero modes.
 These follow easily from their extremal weight property, see \cite{Steinacker:2014eua}.
Since one of these two conditions is always satisfied for any pair of roots $\a,\b$ of $\msu(3)$,
this cubic term vanishes  for the regular zero-modes, so that their interaction potential is
\begin{align}
 V(\phi)
 &= \tr \big( \frac 12 \phi^\a  \big(\Box_X + 2 \Di_{\rm diag}^X \big) \phi_\a 
 + \frac 14[\phi_\a,\phi_\b][\phi^\a,\phi^\b]   \big)  + V_{\rm soft}(\phi) \ .
\end{align}
where the quadratic term vanishes  in the absence of mass terms.
The argument applies also to Higgs modes connecting stacks 
of branes, as long as the $X_\a$ are proportional to $\msu(3)$ generators.
Note that the Higgs potential has similar structure as our starting point \eq{V-soft}.
Although a full analysis of this potential is beyond the scope of this paper, 
it is plausible that the cubic  flux term $V_3(\phi)$ again
induces a non-trivial VEV to some of the Higgs modes, which are 
stabilized by the quartic term.
A deformation of the branes by quantum corrections or mass terms\footnote{Another conceivable mechanism 
is a rotation of the branes, see \cite{Steinacker:2014eua}.} might also
play an important role here.


The above argument for \eq{V-phi-rewrite} to vanish does not apply to the exceptional zero modes.
Among those, the $SU(3)_R$ Goldstone bosons are exactly flat directions, but the 
$\L\in\cW(1,0), \ \L'\in\cW(2,0)$ (or conjugate) modes connecting $\cC[(0,1)]$ with $\cC[(1,0)]$
might lead to non-trivial cubic terms. Again, this needs to be studied in more detail elsewhere.

Finally, we emphasize that even though the Higgs sector consists of many distinct fields 
$\tilde\phi_{\a},  \phi_\a$ etc., there should nevertheless  be one  
lowest Higgs {\em fluctuation} mode around the common minimum,
which is likely  a combination of all the $\tilde\phi_{\a}$ and $\phi_\a$ modes.
Thus the assumption $\tilde\varphi \gg \varphi$ is not in obvious 
conflict with observation.
At higher energies of course, several distinct Higgs modes will necessarily show up.

\section{Numerical results and checks}
\label{sec:numerics}

Since the detailed structure of the various zero modes and their Yukawa couplings
is quite complicated, 
a background consisting of two minimal branes with Higgs and an extra point brane
was implemented in Mathematica.
We consider two branes $\cC[(1,0)] + \cC[(0,1)]$ linked by a Higgs $\phi_{\a}$
and  $\tilde\phi_{\a}$ as in \eq{Higgs-i-structure},
and add a point brane $\cC[0]$ to this configuration. 
We are interested in the Yukawa couplings and the masses of the fermions in this background,
which is determined by the low-energy spectrum and the eigenmodes of the 
Dirac operator $\slashed{D}_{(\rm int)}$ 
acting on spinors 
\begin{align}
  \Psi \ &\in \ \C^8 \otimes \End\big((1,0) + (0,1) + (0,0)\big) \nn\\
  &= \C^8 \otimes \End((1,0) + (0,1)) + \C^8 \otimes(2(1,0) + 2(0,1)) + \C^8 \otimes (0,0)  . 
 \label{full-end-decomp}
\end{align}
Here we note that $(0,0)\cong\C$ and 
\begin{align}
 \Hom((1,0) + (0,1),\C) \cong \Hom(\C,(1,0) + (0,1)) \cong (1,0) + (0,1) .
 \label{link-point-hom}
\end{align}
Due to the MW condition, these two contributions are identified, so that the last term in \eq{full-end-decomp} 
reduces to $(1,0) + (0,1)$; a similar reduction should be applied to all modes.
The lowest eigenvalues and multiplets 
of the Dirac operator 
$\slashed{D}_{(\rm int)}$ on the background $X_\a + \phi_\a + \tilde\phi_\a$
were obtained as a function of the parameters $\varphi_i, \tilde\varphi_i$,
for $r \gg \tilde\varphi \gg \varphi$.
Their $U(1)_i^K$ eigenvalues $\L'$ have also been determined. 
The detailed results are as follows:

\subsection{Fermions linking $\cC[(1,0)] + \cC[(0,1)]$ to a point brane} 
\label{sec:2branes-Higgs-point}

The most interesting sector are the fermionic links \eq{link-point-hom}
of a point brane to $\cC[(1,0)] + \cC[(0,1)]$, which we discuss first. 
They are determined by the Dirac operator $\slashed{D}_{(\rm int)}$ 
acting on spinors 
\begin{align}
 \Psi \in \C^8 \otimes ((1,0) \oplus (0,1))
\end{align}
as in \eq{zeromodes-fermions-L0-1}.
In the absence of any Higgs $\varphi_i = 0 = \tilde\varphi_i$, there are 6+6 
exact zero modes as expected on $\cC[(1,0)] + \cC[(0,1)]$, 
and the non-zero eigenvalues of $\slashed{D}_{(\rm int)}$
are of order $r$.  Switching on $\tilde\varphi_i\equiv\tilde\varphi$ but
leaving $\varphi_i=0$,  six of the would-be zero modes  (``mirror fermions'') acquire 
non-vanishing eigenvalues\footnote{these specific eigenvalues are not hard to understand.} 
$\tilde\varphi(4,-4,2,2,-2,-2)$,
while $3+3$ exact zero modes remain.
The latter  are the ``light fermions'' which constitute the fermions in the standard model,
and one can verify that \eq{chi-Xi-zeromodes} holds\footnote{This holds to an excellent approximation
as long as the background is undeformed, i.e. $r \gg \tilde\varphi$.}, i.e. their chirality $\chi$ is measured by 
$\Xi = \one_3 - \one_{\bar 3}$.
These are indeed modes with $\L'\in\cW(1,2)$.
Switching on also $\varphi_i \equiv \varphi\ll \tilde\varphi$, these light fermions acquire 
eigenvalues approximately given by $\varphi(4,-4,2,2,-2,-2)$.
This precisely confirms  the analysis in the previous sections, which means that the 
low-energy leptons and quarks on the SM brane configurations have indeed the appropriate chiral structure.


\subsection{Fermions within $\cC[(1,0)] + \cC[(0,1)]$} 
\label{sec:2branes-Higgs}

Now consider the fermions on $\cC[(1,0)] + \cC[(0,1)]$, which live in 
\begin{align}
 \Psi \ &\in \ \C^8 \otimes \End((1,0) \oplus (0,1)) \nn\\
  &= \C^8 \otimes \big(2\times(1,1) + (2,0) + (0,2) + (0,1) + (1,0) +  2\times (0,0) \big) \ .
\label{LR-fermions-numeric}
 \end{align}
In the absence of any Higgs $\varphi_i = 0 = \tilde\varphi_i$, we find indeed $52 = 6*6+2*8$ 
exact zero modes, which
are the superpartners of the regular scalar zero modes in this sector.
The $(1,1) + (0,0)$ preserve the branes, while the $(2,0) + (0,1)$
modes connect the two branes. The remaining 
non-zero eigenvalues of $\slashed{D}_{(\rm int)}$ are of order $r$.

 Switching on $\tilde\varphi_i\equiv\tilde\varphi$ but $\varphi_i=0$ leaves $20=8+6+6$ exact zero modes, 
 and 4 low-mass modes of order $\cO(\tilde\varphi^2/r)$. 
 Clearly 8 zero modes arise from the trivial matrix wavefunction 
 $\psi \sim\id_\cH$, 
 which decompose into 6  zero modes with  
 $\L'\in \cW(1,1)$, and two (gaugino) modes with $\L'=0$.
 Six further zero modes have $\L'\in \cW(3,1)$ or $\cW(1,3)$ 
 with $\varXi=\pm 2$,  corresponding to ``mirror'' Higgsinos connecting the branes. 
 The remaining 6 zero modes are a mixture of $\L'\in \cW(2,2)$ 
 and 
 $\L'\in \cW(1,1)$ modes on the branes,
 which are  brane-preserving $\varXi=0$.
 

Besides these 20 zero modes,  the 
4 lowest-mass modes have $\varXi = \pm 2$  and 
$\L'\in\cW(1,2)$ or $\L'\in\cW(2,1)$. Hence these are Higgsino modes connecting the branes.
 

Switching on also $\varphi_i = \varphi\ll \tilde\varphi$, only the 8 trivial zero modes 
$\sim id_\cH$ modes remain, followed by a series of low-mass modes starting with 6 modes 
of order $\cO(\varphi^2/r)$.

We note that \eq{LR-fermions-numeric} also describes the fermions connecting up
and down branes, since the representations are the same.
This therefore covers the entire upper-left
$4\times 4$ block in \eq{fermion-matrix}.

\subsection{$\cC[(1,0)] + \cC[(0,1)] + \cC[0]$ with $\phi_S$}
\label{sec:2+1branes-Higgs}

Now we take the full configuration $\cC[\mu_L] + \cC[\mu_R] + \cC[0]$
with Higgs $\phi_\a,\tilde \phi_\a$ as above, 
organized as of 3+3 modes with $\t$-parity $\t=+1$ as in \eq{Higgs-S-structure} 
\begin{align}
 \phi^{+}_{iS} &\equiv \,  \phi_{\a_i S}^{(0)} = \varphi_{i S}|\mu^{i+1}_R\rangle \langle 0|, \qquad \L'\in\cW(1,2) \nn\\
 \tilde\phi^{+}_{iS} &\equiv \,  \tilde\phi_{\a_i S}^{(0)} = \tilde\varphi_{i S}|0\rangle \langle \mu^{i}_R|, 
   \qquad \L'\in\cW(2,1)
\end{align}
in the basis \eq{Higgs-i-structure}. 
Just like the Higgs $\phi_\a,\tilde\phi_\a$, not all of them need to be switched on.

Consider first the case $\varphi_\a = 0, \tilde\varphi_\a \neq 0$.
If both $\varphi_{i S} = 0 = \tilde\varphi_{i S}$, we have the situation discussed above,
i.e. 20 zero modes on the $\cC[(1,0)] + \cC[(0,1)]$ branes, $2\times 6$ 
massless fermions\footnote{The factor 2 comes from the doubling in \eq{link-point-hom},
which is eliminated by the MW constraint.} between 
$\cC[0]$ and the others, and 8 trivial zero modes on $\cC[0]$.

Switching on $\tilde\varphi \approx\varphi_{i S} \neq 0$ but keeping $\tilde\varphi_{i S} = 0$
gives 16 exact zero modes, while the lowest non-vanishing multiplet 
consists of 4 states with eigenvalue of order $\cO(\frac{\varphi_S\tilde\varphi}{r})$.
8 of these zero modes are easily identified as 
trivial $\psi \sim\id_\cH$ modes. 
The remaining 8 zero modes consist of six 
$\L'\in \cW(3,1)$ or $\cW(1,3)$ 
 with $\varXi=\pm 2$ corresponding to extra Higgsinos connectings the branes,
and two $\L'=0$ modes which preserve the branes. 
Clearly $\varphi_S\neq 0$ gives mass to the  6 modes of $\L'\in \cW(2,2)$ 
 and  $\L'\in \cW(1,1)$ modes on the branes found in section \ref{sec:2branes-Higgs}.
The 4 lowest non-zero modes are essentially $\L'\in\cW(2,1)$ or  $\L'\in\cW(1,2)$ modes connecting 
$\cC[\mu_L]$ and $\cC[\mu_R]$ to $\cC[0]$.


Exchanging the roles of $\tilde\varphi_{i S}$ and $\varphi_{i S}$ 
gives a rather different picture. Switching on $\tilde\varphi_{i S} \neq 0$ but keeping $\varphi_{i S} = 0$
leaves only 8 exact zero modes, and a number of very low but nonzero modes.
The 8 zero modes are again the trivial $\psi \sim\id_\cH$ modes.
The remaining 4 lowest non-trivial modes are found to be 
4 brane-preserving $\L'\in\cW(1,1)$ modes.
 Among the non-zero modes, there is clearly a seesaw-like 
mechanism at work, since the eigenvalues are much smaller than any of the 
$\varphi,\varphi_S$ scales. For example setting $r=10$ and $\tilde\varphi = \varphi_S = 1$
gives $10^{-4}$ as lowest non-trivial eigenvalue.

Finally switching on also $\varphi_i = \varphi\ll \tilde\varphi$
leaves only  8 exact zero modes $\sim id_\cH$, and a number of low eigenvalues, again with a seesaw-like 
mechanism lowering some of the eigenvalues. 
For example setting $r=10$ and $\tilde\varphi = \varphi_S = \tilde\varphi_S=1$
gives $10^{-2}$ as lowest non-trivial eigenvalue.
Again, half of these modes will be eliminated by the MW constraint.

It is interesting to observe that $\cC[(0,1)] + \cC[(1,0)] + \cC[0]$ 
with both Higgs switched on corresponds to the decomposition of the $(7)=(3) + (\obar 3) + (1)$ 
of $G_2$ under $\msu(3)_X$.
There is in fact such a solution of our model, albeit an unstable one. The precise Higgs structure 
and its minima is clearly complicated and will be studied elsewhere.

\subsection{Generic squashed $\cC[\mu]$ branes}

Finally, we briefly discuss the case of generic branes with non-minimal $\mu$.
If the Higgs modes are again realized as links between the extremal weight states 
of the $\cH_{\mu_{L}}$ and $\cH_{\mu_{R}}$, the story goes through with minor modifications.
One important difference is that the masses of the (mirror) fermions will now be larger than the electroweak scale,
due to the enhancement factor $\sqrt{\dim\cH}$ in \eq{higgs-coupling-g}.
This should help to make the present scenario more realistic.
The quark and lepton sector which arises from 
$\Hom(\C,\cH_{\mu_{L,R}})$ is qualitatively the same as in the minimal case, since 
any $\cH_{\mu_{L,R}}$ leads to precisely 3+3 chiral fermionic zero modes.
Hence much of the discussion of this paper is in fact quite generic.
Although the mode decomposition $\End(\cH_{\mu_L},\cH_{\mu_R})$
will be more complicated leading to more Higgs-like multiplets,
the decomposition into 
chiral and mirror sectors should work as in the minimal case.

\section{Summary and discussion}

We have (re-)derived the fermionic and bosonic zero modes which arise on stacks of squashed 
$\cC[\mu]$ brane solutions in $\cN=4$ SYM \cite{Steinacker:2014lma}, deformed by a cubic SUSY-breaking potential corresponding to a 
holomorphic 3-form. These modes are organized in terms of two unbroken global 
gauged $U(1)_{K_i}$ symmetries, which provides a useful tool to understand their interactions.
We use this to start exploring possible symmetry breaking patterns which arise from giving VEV's
to these massless scalar fields (dubbed ``Higgs'' modes), and to study the 
resulting low-energy physics. One important result is that there are 
possible Higgs configurations  which lead to a chiral low-energy theory, in the sense that 
different chiralities of the fermionic (would-be) zero modes couple differently to the 
spontaneously broken massive gauge fields.

To explore the possible implications,
we discuss a  brane configuration which leads to an extension of the standard model, 
correctly reproducing the leptons and quarks with the appropriate coupling to the 
low-energy gauge bosons, {\em assuming} an appropriate Higgs configuration.
This can be viewed as an extension of the MSSM, where each chiral super-multiplet
has an extra mirror copy with the opposite chirality, and acquires a higher (by assumption)
mass from the mirror Higgs. 
This is reminiscent of mirror models \cite{Maalampi:1988va}, 
with the particular feature that  the 
Higgs multiplets also have  mirror partners, which couple only to the mirror fermions.
Thus the light and the mirror sectors communicate only via the common gauge fields,
and through the lowest Higgs  excitation modes which are expected to be a combination of
the different multiplets.
The mirror copies carry different quantum numbers under the $U(1)_{K_i}$
and the opposite $\t$-parity,
and are thereby protected from recombining. Some fields come in different varieties,
and might acquire masses from different Higgs modes.
However due to the complicated Higgs sector, no attempt is made in this paper
to find the minima and to justify the assumed Higgs configuration.

Even if it may seem unlikely that such a scenario could be realistic, 
it is certainly worthwhile to explore the possible scope of these deformed $\cN=4$ models,
given their special status in field theory. 
The most obvious issue seems to be the requirement that the mirror Higgs $\tilde \phi_i$
should give a large mass to the mirror fermions, while it also couples to the $W$ and $Z$
bosons and thereby gives the dominant contribution to their masses.
This means that $\langle\tilde \phi\rangle$ must be at the electroweak scale.
On the other hand, the Yukawa couplings may be large for
large branes (cf. \eq{higgs-coupling-g}), so that the mirror fermions may indeed be much heavier 
than the electroweak scale. In any case, a  more
detailed knowledge of the Higgs sector and its lowest
fluctuations is required before further conclusion can be drawn.

It is  important to stress that although the low-energy spectrum of the squashed brane solutions
is ``mostly'' supersymmetric, there 
are  exceptional scalar zero modes which do not have any fermionic counterpart.
Therefore SUSY is manifestly broken.
One set of such exceptional zero modes 
are the $SU(3)_R$ Goldstone bosons. Two of them are equivalent to gauge transformations and hence unphysical,
and the remaining would disappear in the presence of mass terms; these could also break the
$\Z_3$ family symmetry. 
Since SUSY is  broken, the low-energy action is extracted from the full
underlying deformed $\cN=4$ theory.

There are many  issues which should be addressed in further work. 
The most important problem is to elucidate the Higgs sector for stacks of branes, 
in particular to see if the configurations assumed in this paper can be justified dynamically.
This could be addressed within the weak coupling regime. 
Another natural step is the generalization to non-minimal branes, which
should allow to lift the mirror sector sufficiently high above the electroweak scale. 
Orbifold versions of the model might  eliminate the mirror sector altogether
(cf. \cite{Chatzistavrakidis:2010xi,Bershadsky:1998cb}).  
In the context of string theory, possible ``global'' realizations of analogous brane configurations 
in a suitable flux background should be sought, and the fate of the various $U(1)$'s 
should be clarified.
Furthermore, a dual description in terms of supergravity might help to shed light on  the 
strong coupling regime.
Finally, the considerations in this paper can be carried over immediately to 
the IKKT matrix model \cite{Ishibashi:1996xs} and suitable deformations, 
which reduces to $\cN=4$ SYM on $\R^4_\theta$ \cite{Aoki:1999vr}. 
Then axion-like fields arise \cite{Steinacker:2007dq}, which might also give mass to the 
extra $U(1)_B$ gauge field.
In any case, it is clear that the present type of brane configurations in deformed $\cN=4$ SYM 
provides a remarkably rich basis for further investigations.

\paragraph{Acknowledgements.}

This work is supported by the Austrian Science Fund (FWF) grant P24713.
I would like to thank L. Alvarez-Gaume, N. Arkani-Hamed, P. Anastasopoulos, R. Blumenhagen, A. Kleinschmidt, 
O. Ganor, D. L\"ust, M. Staudacher, 
S. Theisen and G. Zoupanos for useful discussions, and J. Zahn for related collaboration.
I would also like to thank the theory division at CERN and the AEI Golm for hospitality.

\appendix

\section{Appendix A: Relation with $\cN=1^*$}
\label{sec:N=1}

It is interesting to note that the present model can be viewed as a (mass deformation of a) 
supersymmetric $\cN=1^*$ deformation of $\cN=4$ SYM, with the superpotential 
\cite{Karch:1999pv,Argyres:1999xu,Polchinski:2000uf}
\begin{align}
 W = \frac{\sqrt{2}}{g_N}\tr([\Phi_1^+,\Phi_2^+] \Phi_3^- -  m \Phi_3^- \Phi_3^-) \ 
\end{align}
choosing $\Phi_1^+,\Phi_2^+,\Phi_3^-$ as the three chiral superfields. 
This gives the following F-term contribution to the scalar potential 
\begin{align}
 \cV_F &= \frac{\del W}{\del \Phi_1^+}\Big(\frac{\del W}{\del \Phi_1^+}\Big)^* 
  + \frac{\del W}{\del \Phi_2^+}\Big(\frac{\del W}{\del \Phi_2^+}\Big)^* 
  +\frac{\del W}{\del \Phi_3^-}\Big(\frac{\del W}{\del \Phi_3^-}\Big)^*  \nn\\
  &= \frac 2{g_N^2}\tr \Big([\Phi_2^+,\Phi_3^-] [\Phi_3^+,\Phi_2^-] + [\Phi_1^+,\Phi_3^-] [\Phi_3^+,\Phi_1^-] 
   + ([\Phi_1^+,\Phi_2^+] - 2m \Phi_3^-) ([\Phi_2^-,\Phi_1^-] - 2m \Phi_3^+) \Big) . \nn
\end{align}
Writing $\Phi = mX$ and adding the D term, the full potential takes the form 
\begin{align}
 \cV &= \frac{m^4}{g_N^2}V(X), \nn\\
 V(X)  &= - \tr\big([X_i^+,X_j^+][X_i^-,X_j^-] - \frac 12 [X_i^+,X_i^-][X_j^+,X_j^-]\big) \nn\\
   &\qquad + 4 \tr\big(-[X_1^+,X_2^+] X_3^+ -[X_2^-,X_1^-] X_3^- +2 X_3^+ X_3^- \big) \nn\\
    &= -\frac 1{4} \sum_{\a\neq \b} \tr [X^\a,X^\b][X_\a,X_\b] 
    + 4\tr\big(-[X_1^+,X_2^+] X_3^+ - [X_2^-,X_1^-] X_3^- +2 X_3^+ X_3^-\big)  .
\end{align}
This is precisely the  potential in \eq{V-soft}, with
\begin{align}
 M_3^2 = 2, \quad M_1=M_2=0  .
\end{align}
Then the global R-symmetry is reduced to $SU(2) \times U(1)$.
However, this value of $M_3^2$ is too large for \eq{eom-general} to admit squashed brane solutions;
these only exist for  $M_3^2 < \frac 43$. 
Thus in the range of $M_i^2$ of interest here, 
the model is not supersymmetric, but can  be viewed as a mass deformation 
of the supersymmetric $\cN=1^*$ model, deformed by a negative mass term $\d M_3^2$. 
This might still be useful to obtain insights into the strong coupling regime.

Adding also mass terms $M_1^2$ and $M_2^2$, the global symmetry is broken to $U(1)\times U(1)$.
Then there are no physical Goldstone bosons on the squashed brane backgrounds, 
since both are equivalent to a gauge transformation.

\section{Appendix B: Exceptional modes as Goldstone bosons}
\label{sec:goldstone}

Here we show that the 6 exceptional zero modes arising from $(1,1) \subset \End(\cH)$
are nothing but the 6 Goldstones arising from $SU(3)_R$ minus the two $U(1)$,
which  are gauged hence eaten by the massive gauge bosons.
To start, recall from \cite{Steinacker:2014lma} that the exceptional zero modes from 
$(1,1) \subset \End(\cH)$
correspond to the extremal weight states with $\L'\in\cW(3,0)$ and $\L'\in\cW(0,3)$.
Denoting the  background as
\begin{align}
 X \sim \l^\a T_\a =  \l_{-\a_1} T_{\a_1} + \l_{\a_3} T_{-\a_3} + ... ,
\end{align}
these arise from the anti-symmetric tensor product $(3,0) \subset (1,1)\otimes(1,1)$,
e.g. 
\begin{align}
 \phi = \l^\a \phi_\a = \l_{-\a_1} T_{\a_3} - \l_{\a_3} T_{-\a_1} 
\end{align}
etc.
Thus 
\begin{align}
 \d T_{\a_1} &= \e T_{\a_3} \nn\\
 \d T_{-\a_3} &= - \e T_{-\a_1}
\end{align}
for any $\e\in\C$, with conjugate mode 
\begin{align}
 \d T_{-\a_1} &= \bar\e T_{-\a_3} \nn\\
 \d T_{\a_3} &= - \bar\e T_{\a_1} .
\end{align}
Thus 
\begin{align}
 \phi_\a = \begin{pmatrix}
            0 &0 &\e\\
            0 & 0 & 0 \\
            -\bar\e & 0 & 0
           \end{pmatrix}_{\a\b} T_\b
\end{align}
etc., which
generates precisely the $SU(3)/(U(1)\times U(1))$ R-symmetry.


\begin{thebibliography}{99}

  
  \bibitem{DuboisViolette:1989at} 
  M.~Dubois-Violette, J.~Madore and R.~Kerner,
  ``Gauge Bosons in a Noncommutative Geometry,''
  Phys.\ Lett.\ B {\bf 217}, 485 (1989).
  

\bibitem{Myers:1999ps} 
  R.~C.~Myers,
  ``Dielectric branes,''
  JHEP {\bf 9912}, 022 (1999)
  [hep-th/9910053].
  
  
  \bibitem{Polchinski:2000uf} 
  J.~Polchinski and M.~J.~Strassler,
  ``The String dual of a confining four-dimensional gauge theory,''
  hep-th/0003136.
  
  
\bibitem{Iso:2001mg} 
  S.~Iso, Y.~Kimura, K.~Tanaka and K.~Wakatsuki,
  ``Noncommutative gauge theory on fuzzy sphere from matrix model,''
  Nucl.\ Phys.\ B {\bf 604}, 121 (2001)
  [hep-th/0101102].

  
 \bibitem{Berenstein:2002jq} 
  D.~E.~Berenstein, J.~M.~Maldacena and H.~S.~Nastase,
  ``Strings in flat space and pp waves from N=4 superYang-Mills,''
  JHEP {\bf 0204}, 013 (2002)
  [hep-th/0202021].

  
\bibitem{Andrews:2006aw} 
  R.~P.~Andrews and N.~Dorey,
  ``Deconstruction of the Maldacena-Nunez compactification,''
  Nucl.\ Phys.\ B {\bf 751}, 304 (2006)
  [hep-th/0601098];
  R.~P.~Andrews and N.~Dorey,
  ``Spherical deconstruction,''
  Phys.\ Lett.\ B {\bf 631}, 74 (2005)
  [hep-th/0505107].
  
  
 \bibitem{Aschieri:2006uw}
  P.~Aschieri, T.~Grammatikopoulos, H.~Steinacker and G.~Zoupanos,
  ``Dynamical generation of fuzzy extra dimensions, dimensional reduction and symmetry breaking,''
  JHEP {\bf 0609} (2006) 026
  [hep-th/0606021].
  
  \bibitem{Steinacker:2007ay}
  H.~Steinacker and G.~Zoupanos,
  ``Fermions on spontaneously generated spherical extra dimensions,''
  JHEP {\bf 0709} (2007) 017
  [arXiv:0706.0398 [hep-th]]
  
  

\bibitem{Steinacker:2014lma} 
  H.~C.~Steinacker and J.~Zahn,
  ``Self-intersecting fuzzy extra dimensions from squashed coadjoint orbits in $ \mathcal{N}=4 $ SYM and matrix models,''
  JHEP {\bf 1502}, 027 (2015)
  [arXiv:1409.1440 [hep-th]].
  
  
\bibitem{Steinacker:2014eua} 
  H.~C.~Steinacker,
  ``Spinning squashed extra dimensions and chiral gauge theory from $\mathcal{N}=4$ SYM,''
  arXiv:1411.3139 [hep-th].

  
  
 \bibitem{Steinacker:2014fja} 
  H.~C.~Steinacker and J.~Zahn,
  ``An extended standard model and its Higgs geometry from the matrix model,''
  PTEP {\bf 2014}, no. 8, 083B03 (2014)
  [arXiv:1401.2020 [hep-th]].
  

  
   \bibitem{Aoki:2014cya} 
  H.~Aoki, J.~Nishimura and A.~Tsuchiya,
  ``Realizing three generations of the Standard Model fermions in the type IIB matrix model,''
  JHEP {\bf 1405}, 131 (2014)
  [arXiv:1401.7848 [hep-th]];
   J.~Nishimura and A.~Tsuchiya,
  ``Realizing chiral fermions in the type IIB matrix model at finite N,''
  JHEP {\bf 1312}, 002 (2013)
  
  
\bibitem{Aldazabal:2000cn}
 G.~Aldazabal, S.~Franco, L.~E.~Ibanez, R.~Rabadan and A.~M.~Uranga,
  ``Intersecting brane worlds,''
  JHEP {\bf 0102} (2001) 047
  [hep-ph/0011132].
  I.~Antoniadis, E.~Kiritsis, J.~Rizos, T.~N.~Tomaras,
  ``D-branes and the standard model,''
  Nucl.\ Phys.\  {\bf B660 } (2003)  81-115.
  [hep-th/0210263];
  C.~Kokorelis,
  ``Exact standard model structures from intersecting D5-branes,''
  Nucl.\ Phys.\ B {\bf 677}, 115 (2004)
  [hep-th/0207234].
  R.~Blumenhagen, M.~Cvetic, P.~Langacker, G.~Shiu,
  ``Toward realistic intersecting D-brane models,''
  Ann.\ Rev.\ Nucl.\ Part.\ Sci.\  {\bf 55 } (2005)  71-139.
  [hep-th/0502005].
   
  \bibitem{Manousselis:2004xd} 
  P.~Manousselis and G.~Zoupanos,
  ``Dimensional reduction of ten-dimensional supersymmetric gauge theories in the N=1, D=4 superfield formalism,''
  JHEP {\bf 0411}, 025 (2004)
  [hep-ph/0406207].
  
  
\bibitem{Karch:1999pv} 
  A.~Karch, D.~Lust and A.~Miemiec,
  ``New N=1 superconformal field theories and their supergravity description,''
  Phys.\ Lett.\ B {\bf 454}, 265 (1999)
  [hep-th/9901041].
  
  
  \bibitem{Argyres:1999xu} 
  P.~C.~Argyres, K.~A.~Intriligator, R.~G.~Leigh and M.~J.~Strassler,
  ``On inherited duality in N=1 d = 4 supersymmetric gauge theories,''
  JHEP {\bf 0004}, 029 (2000)
  [hep-th/9910250].
  
  
  \bibitem{Vafa:1994tf} 
  C.~Vafa and E.~Witten,
  ``A Strong coupling test of S duality,''
  Nucl.\ Phys.\ B {\bf 431}, 3 (1994)
  [hep-th/9408074].
  
  
\bibitem{Tseytlin:1999tp} 
  A.~A.~Tseytlin and K.~Zarembo,
  ``Magnetic interactions of D-branes and Wess-Zumino terms in superYang-Mills effective actions,''
  Phys.\ Lett.\ B {\bf 474}, 95 (2000)
  [hep-th/9911246]; 
  D.~V.~Belyaev and I.~B.~Samsonov,
  ``Wess-Zumino term in the N=4 SYM effective action revisited,''
  JHEP {\bf 1104}, 112 (2011)
  [arXiv:1103.5070 [hep-th]].
  
  
 \bibitem{Andronache:2015sxa} 
  S.~Andronache and H.~C.~Steinacker,
  ``The squashed fuzzy sphere, fuzzy strings and the Landau problem,''
  arXiv:1503.03625 [hep-th].
  
  
  \bibitem{Chatzistavrakidis:2011gs}
  A.~Chatzistavrakidis, H.~Steinacker and G.~Zoupanos,
  ``Intersecting branes and a standard model realization in matrix models,''
  JHEP {\bf 1109} (2011) 115
  [arXiv:1107.0265 [hep-th]].

  
  \bibitem{Blaschke:2011qu} 
  D.~N.~Blaschke and H.~Steinacker,
  ``On the 1-loop effective action for the IKKT model and non-commutative branes,''
  JHEP {\bf 1110}, 120 (2011)
  [arXiv:1109.3097 [hep-th]].
  
  \bibitem{Chepelev:1997av} 
  I.~Chepelev and A.~A.~Tseytlin,
  ``Interactions of type IIB D-branes from D instanton matrix model,''
  Nucl.\ Phys.\ B {\bf 511}, 629 (1998)
  [hep-th/9705120].
  
  
\bibitem{Ishibashi:1996xs}
  N.~Ishibashi, H.~Kawai, Y.~Kitazawa, A.~Tsuchiya,
  ``A Large N reduced model as superstring,''
  Nucl.\ Phys.\  {\bf B498 } (1997)  467-491.
  [hep-th/9612115].
  
  
  \bibitem{Steinacker:2010rh} 
  H.~Steinacker,
  ``Emergent Geometry and Gravity from Matrix Models: an Introduction,''
  Class.\ Quant.\ Grav.\  {\bf 27}, 133001 (2010)
  [arXiv:1003.4134 [hep-th]].
  
  
  \bibitem{Camara:2003ku} 
  P.~G.~Camara, L.~E.~Ibanez and A.~M.~Uranga,
  ``Flux induced SUSY breaking soft terms,''
  Nucl.\ Phys.\ B {\bf 689}, 195 (2004)
  [hep-th/0311241].
 
 \bibitem{Grana:2003ek} 
  M.~Grana, T.~W.~Grimm, H.~Jockers and J.~Louis,
  ``Soft supersymmetry breaking in Calabi-Yau orientifolds with D-branes and fluxes,''
  Nucl.\ Phys.\ B {\bf 690}, 21 (2004)
  [hep-th/0312232].
  
\bibitem{Morelli:2009ev} 
  S.~Morelli,
  ``St\"uckelberg Axions and Anomalous Abelian Extensions of the Standard Model'',
  PhD thesis Salento,
  arXiv:0907.3877 [hep-ph].

\bibitem{Anastasopoulos:2006cz} 
  P.~Anastasopoulos, M.~Bianchi, E.~Dudas and E.~Kiritsis,
  ``Anomalies, anomalous U(1)'s and generalized Chern-Simons terms,''
  JHEP {\bf 0611}, 057 (2006)
  [hep-th/0605225].
  
  \bibitem{Coriano':2005js} 
  C.~Coriano, N.~Irges and E.~Kiritsis,
  ``On the effective theory of low scale orientifold string vacua,''
  Nucl.\ Phys.\ B {\bf 746}, 77 (2006)
  [hep-ph/0510332].


\bibitem{Coriano:2006xh} 
  C.~Coriano and N.~Irges,
  ``Windows over a New Low Energy Axion,''
  Phys.\ Lett.\ B {\bf 651}, 298 (2007)
  [hep-ph/0612140].
  
  \bibitem{Coriano:2007fw} 
  C.~Coriano, N.~Irges and S.~Morelli,
  ``Stuckelberg axions and the effective action of anomalous Abelian models. 1. A Unitarity analysis of the Higgs-axion mixing,''
  JHEP {\bf 0707}, 008 (2007)
  [hep-ph/0701010].
  
  
  
  \bibitem{Kors:2005uz} 
  B.~Kors and P.~Nath,
  ``Aspects of the Stueckelberg extension,''
  JHEP {\bf 0507}, 069 (2005)
  [hep-ph/0503208].
  
  \bibitem{Steinacker:2007dq} 
  H.~Steinacker,
  ``Emergent Gravity from Noncommutative Gauge Theory,''
  JHEP {\bf 0712}, 049 (2007)
  [arXiv:0708.2426 [hep-th]]; 
  H.~Steinacker,
  ``Covariant Field Equations, Gauge Fields and Conservation Laws from Yang-Mills Matrix Models,''
  JHEP {\bf 0902}, 044 (2009)
  [arXiv:0812.3761 [hep-th]].
  
\bibitem{Grosse:2010zq}
  H.~Grosse, F.~Lizzi, H.~Steinacker,
  ``Noncommutative gauge theory and symmetry breaking in matrix models,''
  Phys.\ Rev.\  {\bf D81 } (2010)  085034.
  [arXiv:1001.2703 [hep-th]].
  
  
\bibitem{Maalampi:1988va} 
  J.~Maalampi and M.~Roos,
  ``Physics of Mirror Fermions,''
  Phys.\ Rept.\  {\bf 186}, 53 (1990);
   T.~Ibrahim and P.~Nath,
  ``An MSSM Extension with a Mirror Fourth Generation, Neutrino Magnetic Moments and LHC Signatures,''
  Phys.\ Rev.\ D {\bf 78}, 075013 (2008)
  [arXiv:0806.3880 [hep-ph]].
  
  
\bibitem{Chatzistavrakidis:2010xi} 
  A.~Chatzistavrakidis, H.~Steinacker and G.~Zoupanos,
  ``Orbifolds, fuzzy spheres and chiral fermions,''
  JHEP {\bf 1005}, 100 (2010)
  [arXiv:1002.2606 [hep-th]].
  
  \bibitem{Bershadsky:1998cb} 
  M.~Bershadsky and A.~Johansen,
  ``Large N limit of orbifold field theories,''
  Nucl.\ Phys.\ B {\bf 536}, 141 (1998)
  [hep-th/9803249].
  
  
\bibitem{Aoki:1999vr}
  H.~Aoki, N.~Ishibashi, S.~Iso, H.~Kawai, Y.~Kitazawa and T.~Tada,
  ``Noncommutative Yang-Mills in IIB matrix model,''
  Nucl.\ Phys.\ B {\bf 565} (2000) 176
  [hep-th/9908141];
  
  
  
  
  
  
  
  
  
  
  
\end{thebibliography}
\end{document}